\documentclass[twocolumn]{aa}

\usepackage{graphicx}
\usepackage{txfonts}
\usepackage{float}
\usepackage{natbib}
\bibpunct{(}{)}{;}{a}{}{,} 
\usepackage{xcolor}

\usepackage{hyperref}
\hypersetup{
    colorlinks=true,
    allcolors=blue
}

\begin{document}

   \title{Recovery of coronal dimmings}


   \author{G. M. Ronca
          \inst{1} \inst{2}
          \and
          G. Chikunova\inst{2}
          \and
          K. Dissauer \inst{3}
          \and
          T. Podladchikova\inst{2}
          \and
          A. M. Veronig \inst{4} \inst{5}
          }

   \institute{Politecnico di Milano, Piazza Leonardo da Vinci, 32, 20133, Milano, taly
        \and
        Skolkovo Institute of Science and Technology, Bolshoy Buolevard 30, bld. 1, Moscow 121205, Russia
         \and
            NorthWest Research Associates, 380 Mitchell Lane, Boulder, CO 80301, USA
        \and
            University of Graz, Institute of Physics, Universitätsplatz 5, 8010 Graz, Austria
        \and
            University of Graz, Kanzelhöhe Observatory for Solar and Environmental Research, Kanzelhöhe 19, 9521 Treffen, Austria
             }

   \date{Paper in prep.}

  \abstract
   {Coronal dimmings are regions of reduced emission in the lower corona observed in the wake of coronal mass ejections (CMEs), representing their footprints. Studying the lifetime evolution of coronal dimmings helps us to better understand the recovery and replenishment of the corona after large-scale eruptions.}
   {We study the recovery of dimmings on different spatial scales to enhance our understanding of the replenishment and dynamics of the corona after CMEs.}
   {In order to investigate the long-term evolution of coronal dimming and its recovery, we propose two approaches that focus on both the global and the local evolution of dimming regions: the fixed mask approach and the pixel boxes approach. We present four case studies (September 6, 2011; March 7, 2012; June 14, 2012; and March 8, 2019) in which a coronal dimming is associated with a flare/CME eruption.
   We analyzed each event with the same methodology, using extreme-ultraviolet filtergrams from the Solar Dynamics Observatory's Atmospheric Imaging Assembly (SDO/AIA) and Solar TErrestrial RElations Observatory's Extreme UltraViolet Imager (STEREO/EUVI) instruments. We identified the dimming region by image segmentation, then restricted the analysis to a specific portion of the dimming and tracked the time evolution of the dimming brightness and area. In addition, we study the behavior of small subregions inside the dimming area, of about 3x3 pixels, to compare the recovery in different regions of the dimming.}
   {Three out of the four cases show a complete recovery 24 hours after the flare/CME eruption. The primary recovery mechanism identified in the observations is the expansion of coronal loops into the dimming region. The recovery of the brightness follows a two-step trend, with a steeper and quicker segment followed by a slower one. In addition, some parts of the dimming, which may be core dimmings, are still present at the end of the analysis time and do not recover within 3 days, whereas the peripheral regions (secondary dimmings) show a full recovery.}
   {The high temporal and spatial resolution of SDO/AIA observations combined with multi-view data of the STEREO/EUV instrument reveal high-situated coronal loops expanding after CME eruptions, which cover dimming regions and gradually increase their intensity. Our developed approaches enable the analysis of dimmings alongside these bright structures, revealing different timescales of recovery for core and secondary or twin dimming regions. Combined with magnetic field modeling, these methods lay the foundation for further systematic analysis of dimming recovery and enhance the knowledge gained from already-analyzed events.}

   \keywords{ Sun: corona -- Sun: coronal mass ejections (CMEs) -- Sun: flares }
   
\authorrunning{Ronca et al.}
\maketitle
%

\section{Introduction}
Among the phenomena that occur on the Sun and that may affect Earth and the Near-Earth environment, coronal mass ejections (CMEs) are the most powerful and violent. Coronal mass ejections are huge clouds of magnetized plasma expelled from the Sun into interplanetary space with speeds in the range of 100 km/s up to >3000 km/s \citep{2009EM&P..104..295G, webb2012coronal} and, as a result, large quantities of mass (billions of tons of coronal material) and magnetic flux are ejected into the heliosphere. 

Earth-directed CMEs are the main drivers of severe space weather effects. 
However, they are the most difficult to detect and parameterize because of projection effects that lead to distortions of the CME structures in images \citep{burkepile2004role}. 
Hence, it is important to also study associated phenomena that are well observed in Earth-directed CMEs, such as coronal dimmings.

Coronal dimmings are regions in the solar corona where a sudden CME-associated decrease in intensity in the soft X-ray (SXR) and extreme ultraviolet (EUV) emission is observed \citep{sterling1997yohkoh, thompson1998soho,dissauer2019statistics}. It has been demonstrated that dimmings are closely associated with fast CMEs \citep{reinard2009relationship, dissauer2018statistics} and are also considered reliable indicators of Earth-directed CMEs, also known as halo CMEs \citep{thompson2000coronal}. 
In the literature, there is a distinction between two different types of dimming: the core dimmings and the secondary dimming regions.
The so-called ``core'' dimmings \citep{mandrini2007cme} are believed to mark the footprint of the erupting flux rope \citep[e.g.,][]{sterling1997yohkoh,webb2000origin, zarro1999soho}. 
Core dimming regions are defined as stationary and localized regions that are close to the eruption site and rooted in opposite magnetic polarity regions.
Secondary dimmings are more shallow and diffuse. They can develop in all directions, either in a symmetrical way or in narrow and extended structures, while reaching significant distances from the eruption site \citep{chertok2005large, attrill2006using}.
The intensity depletion in the solar corona is interpreted to be the result of the plasma escaping and expanding along the magnetic field lines opening during the eruption \citep[e.g.,][]{hudson1996coronal, sterling1997yohkoh, thompson2000coronal}. 
Hence, coronal dimmings are studied to obtain information about the early stages of a CME eruption as well as relevant CME properties, like mass, velocity, and acceleration \citep{harrison2000spectroscopic, harrison2003coronal, aschwanden2009first, aschwanden2016global, mason2016relationship, krista2017statistical, dissauer2018statistics, dissauer2019statistics}.  Recent studies indicate that the dimming expansion can be also used to estimate the early CME propagation direction
\citep{2023Chikunova, Jain2024}. Studies of the lifetime of coronal dimmings and their recovery phase can provide information about the evolution of the CME as well as the replenishment of the corona post-eruption \citep{kahler2001origin, attrill2006using, vanninathan2018plasma}. 
In the related literature, the term ``recovery'' usually indicates the return to (or exceeding of) the pre-dimming intensity \citep{attrill2008recovery, reinard2009relationship}. 
A central goal of this research is to understand how the dimming, which marks the magnetic footprint of a CME rooted in the solar atmosphere, could maintain its relationship to the CME while it expands out into the interplanetary space until reaching Earth \citep{kahler2001origin}. 

The analysis of the recovery of coronal dimmings and the investigation of their duration aims to understand how and when the change back from the open magnetic field to a closed configuration after an eruption occurs. 
The study by \cite{kahler2001origin} proposes two recovery mechanisms: shrinkage of the dimming boundaries inward, without internal brightening (implying no magnetic reconnection), and arcade brightening (indicating magnetic reconnection within the dimming region). 
Analyzing 19 dimming events with the Yohkoh Soft X-ray Telescope (SXT), they found that the dimmings disappeared due to boundary shrinkage and had lifetimes ranging from 5 to 48 hours. This is much shorter than the 3 to 5 days needed for interplanetary CMEs (ICMEs) and magnetic clouds (MCs) to reach Earth, suggesting a disconnect between dimmings and MCs. This contrasts with the theory that coronal dimmings are the footpoints of MCs at 1 AU, as was suggested by \cite{sterling1997yohkoh, webb2000relationship, webb2000origin}.


\cite{attrill2008recovery} studied in detail the mechanism behind coronal dimming disappearance before the associated MC reaches 1 AU by analyzing three case studies of dimming events in the 195 \AA\ filter of the EUV Imaging Telescope (EIT) on board the Solar and Heliospheric Observatory (SOHO) over approximately two days. 
They identified a long-living ``core'' part within the dimming that did not fully recover, unlike the peripheral area; that is, secondary dimming. The average intensity of the dimming, excluding the core, recovered to pre-eruption levels. 
They observed that dimmings recover by shrinking in a fragmentary and inhomogeneous manner, in agreement with \cite{kahler2001origin}. 
However, they showed that there is a progressive increase in intensity within the dimmings, a detail possibly missed in earlier studies due to lower resolution or lower sensitivity in SXRs compared to the EUV.
\cite{attrill2008recovery} estimated the time for the solar corona to recover to pre-event intensity, noting that this did not align with the MC arrival at 1 AU in two out of three cases studied. 
They proposed that the recovery mechanism involves interchange reconnection between the open magnetic field of the dimming and small coronal loops or emerging flux bipoles, dispersing the open field into the surrounding quiet Sun rather than closing the global coronal magnetic field. 
At the boundary of the dimming, bright loops or emerging flux could play a role in diminishing the dimming size and replenishing the flux from the inside, by bringing closed loops into the open magnetic field of the dimming and making the process at the periphery faster than internal recovery.

An interesting methodology for investigating different parts of the secondary dimming was suggested by  \cite{attrill2010revealing}.  
The authors used base difference and original images in different wavelengths from Hinode/EUV imaging spectrometer (EIS) and SOHO/EIT data to understand the recovery process of two dimmings that occurred on December 13 and 14, 2006. They identified four localized regions per event with different sizes and observed light curves of the average intensity within each region. 
The selected regions showed a gradual recovery process, with a substantial recovery within two hours from the dimming onset. 
Despite noticing an up-flowing mass supply from beneath the dimming region, in accordance with \cite{jin2009coronal}, the authors claimed that the reestablishment of overlying coronal loops plays a major role in the recovery process. 
Overall, the recovery process of the analyzed secondary dimmings is similar to the formation of AR flare ribbons and flare loops, but the footpoints are more scattered \citep{attrill2010revealing}.

\cite{reinard2008coronal} conducted the first statistical analysis of dimming recovery time, examining 96 dimming events associated with CMEs. They used area and brightness curves to track dimming evolution. ``Rise time'' is defined as the interval from the start of dimming to its peak (maximum area), while "recovery time" is the interval from the peak to the end. Dimming duration ranged from 1 to 19 hours, with an average value of 8 hours. Most dimmings did not recover to pre-eruption values due to darkening effects from image de-rotation.
The recovery curve showed either a single slope or a two-part slope. About $76\%$ of brightness and $72\%$ of area time series had a single slope. 
In two-part recoveries, the first segment mirrored the slope during the rise phase, followed by a more gradual recovery, suggesting a faster initial plasma inflow followed by a slower one. 
The similarity in slopes between the initial recovery and the dimming rise phase may indicate a similar mechanism in plasma removal and initial restoration.

Another statistical study conducted by \cite{krista2017statistical} focused on 115 dimming systems identified as CME footpoints, which were detected with the Coronal Dimming Tracker \citep[CoDiT, ][]{krista2012study} in direct 193 \AA\ Atmospheric Imaging Assembly \citep[AIA;][]{lemen2012atmospheric} observations. Of these, only 64 dimmings were associated with CMEs, 41 with flares, and the rest occurred in quiet Sun regions. The study delved into various properties of dimmings, including their average lifetime, which was found to be 9 hours, with an average rise time of 3 hours and a recovery time of 6 hours. 
They noted that larger dimmings exhibited longer duration, suggesting a prolonged process in ``closing down'' large open magnetic regions. Moreover, smaller dimmings tended to experience a higher magnetic flux imbalance during their growth phase compared to larger dimmings.

Using a similar methodology, \cite{krista2022study} examined 53 coronal dimmings detected from 195 \AA\ Extreme UltraViolet Imager \citep[EUVI; ][]{wulser2004euvi} data from the twin Solar TErrestrial RElations Observatory \citep[STEREO; ][]{kaiser2008stereo} satellites, STEREO-A and STEREO-B, from December 2010 to August 2011.
The mean lifetime of dimmings was observed to be 11 hours, with a rise time of 2 hours and a recovery time of 9 hours. Their analysis revealed correlations between flare duration and various dimming characteristics, such as total intensity change, pre-eruption intensity, and lifetime, suggesting that longer flare duration corresponds to longer reconnection and mass evacuation processes. Additionally, they found moderate to strong correlations between dimming area and the rise, recovery, and total lifetime, suggesting that larger dimmings take longer to form and dissipate, while long-lived dimmings exhibit increased brightness pre- and post-eruption.

Our analysis fills in this context and seeks to derive for selected events the maximum time ranges over which we can still observe signatures of coronal dimmings associated with a CME.
This duration may contain important information on the timescales of the replenishment of the corona after an eruption has occurred and/or how long the legs of a CME are still connected to the Sun \citep{vanninathan2018plasma}.

However, a problem that has been encountered by all the authors of previous studies is the distortion effect that affects the images when they are rotated to a common reference time to compensate for the solar differential rotation, which introduces uncertainties in the computations of dimming recovery times.
The new approach suggested in this work aims at minimizing the effects of the de-rotation of the images while monitoring the evolution of the solar corona and extracting relevant parameters for the dimming analysis, like instantaneous brightness and area, using a fixed mask approach. 
In addition, a further step is taken in the observation of different parts of the same dimming by suggesting an approach to the observation and distinction between core and secondary dimmings, which may be useful for further understanding the dimming behavior and to obtain more data about the CME development while the flux rope expands out into interplanetary space. In a similar way to what is done by \cite{attrill2010revealing}, we restrict the analysis to a small portion of the image, but by choosing a well-defined and fixed size it is possible to observe the dynamics within the desired regions with a uniform approach.

\section{Data and event overview}
\label{data explaination}

The AIA instrument onboard the Solar Dynamics Observatory \citep[SDO;][]{pesnell2012solar} regularly observes the Sun in seven EUV filters with a cadence as high as 12 seconds, sampling plasma over a temperature range from approximately 50,000 K to 10 MK. In this study, we use filtergrams of the 211 \AA\ passband (with a peak formation temperature of $\log T = 6.3$) because it enables to visualize the magnetically active regions in the Sun's corona. 
In addition, data from the twin STEREO spacecraft are used. These satellites are placed on heliocentric orbits, one ahead of Earth in its orbit (STEREO-A), the other trailing behind (STEREO-B). However, no more data have been available from STEREO-B since September 2016. In this study, we use filtergrams from the Extreme Ultraviolet Imager (EUVI) instrument of the Sun-Earth Connection Coronal and Heliospheric Investigation (SECCHI) package \citep{howard2008sun} in the 195 \AA\ passband, which allows us to observe the solar corona in a similar way to what is done by SDO/AIA. The STEREO satellite that has better visibility of the Active Region (AR) of interest is chosen. 
The events under study occur during quadrature periods of the satellites and coronal dimmings are observed against the solar disk with SDO/AIA and from above the limb with STEREO/EUVI.

The analysis is performed in Python. In particular, this research used version 4.0.0 of the SunPy open-source software package \citep{sunpy_community2020}.
All AIA maps are checked for constant exposure time, then calibrated \citep[from aiapy.calibrate package][]{barnes2020aiapy} 
 and resampled to $2048\times2048$ pixels. 
 
We study the evolution of coronal dimmings associated with four different flare/CME events. For each event, the SDO/AIA time series cover 72 hours, starting 30 minutes before the associated flare, with a cadence of 5 minutes for the first 24 hours and 30 minutes for the next 48 hours (except for September 06, 2011, where the time cadence is 5 minutes for an overall time span of 48 hours). 
The STEREO/EUVI data have a cadence of 5 min for the entire time interval. 
For each event, we identify the start time, peak time, and end time of the flare associated with a CME and check whether other phenomena occur before the end of the 72-hour time interval.
The flare/CME events under study are the following:

\begin{itemize}
    \item September 6, 2011. 
    A first X2.1-class flare (N14 W18), associated with a halo CME of speed 990 km/s (see Tab.\ref{tab: table-review} for references), is produced by NOAA AR 11283 on September 6 at 22:20 UT.
    A second X1.8-class flare on September 7 at 22:38 UT and an M6.7-class flare on September 8 at 15:46 UT originate from the same AR. The latter events are associated with a partial halo CME (angular width of 167°) and a CME of angular width 37°, respectively. The main evolution phase of the coronal dimmings was studied in \cite{dissauer2018statistics} and \cite{vanninathan2018plasma}, who focused on the plasma parameters. The magnetic topology of the CME and associated dimming was studied in \cite{prasad2020magnetohydrodynamic}.
    
    \item March 7, 2012. 
    A X5.4-class flare  (N18 E31) originates from NOAA AR 11429 on March 7 at 00:24 UT followed by a X1.3-class flare at 01:14 UT. Both events are associated with fast halo CMEs, although the second flare was not visible due to being occulted by the first, since the two events occurred in close succession \citep{ajello2014impulsive}. The speed of the first CME was measured at 2680 km/s \footnote{\url{https://cdaw.gsfc.nasa.gov/CME_list/}}. 
    The coronal dimming has been analyzed in \cite{dissauer2018statistics} and \cite{vanninathan2018plasma}. This is the strongest event in the Sun-as-a-star context, which was also examined by \cite{veronig2021detection} in AIA imaging and EVE Sun-as-a-star observations. 
    For this study, we focus on the dimming evolution following the first X5.4 flare/CME on March 7. Notably, the analyzed time range also includes an M6.3-class flare originating on March 9 at 03:58 UT, also associated with a fast halo CME.

    \item June 14, 2012. 
    A long-duration M1.9-class flare (S17 E06), associated with a halo CME of speed 990 km/s \footnotemark[\value{footnote}], originates on June 14 at 14:35 UT from NOAA AR 11504. This event is an example of one occurring in a highly complex AR, considering its surrounding coronal loops and nearby structures.
    
    \item March 8, 2019. 
    A double-peaked C1.3-class flare  (N11 W04) occurs on March 8 at 03:07 UT (the second peak's time is 03:47 UT) and a B6.2-class flare occurs on March 9 at 12:26 UT. They originated from active region AR 12734 and each of them is associated with a CME of angular width 74° and 71°, respectively. The speed of the first CME is 290 km/s \footnotemark[\value{footnote}]. This event occurred during the solar minimum and presents a simpler magnetic configuration (Hale class $\beta$) compared to the other events (Hale class $>\beta\gamma$) that all occurred during the solar maximum of solar cycle 24. The early evolution of the dimming associated with this case was investigated in detail by \citet{dumbovic20212019}. From now on, we shall refer to the March 8, 2019, case as the International Women's Day case, as already used in the literature, to easily distinguish it from the March 7, 2012, event.
    
\end{itemize}

\begin{table*}[h]
\caption{\label{tab: table-review}Event overview.}
\centering
\begin{tabular}{l c c c c c c c c c c } 
        \hline \hline
        Date & NOAA & Flare & Flare & \multicolumn{3}{c}{Time (UT)} & Flare & STEREO & Separation & CME\\ 
        
        ~   & AR & Location & Class &   ~   &   ~  &  ~   & duration & satellite & angle of STEREO & speed \\
        ~   &  ~ &  ~   &  ~  & Start & Peak & End  & [min] & & and SDO [°] & [km/s] \tablefootmark{a} \\
        \hline \hline
        \noalign{\smallskip}
        \noalign{\smallskip}
        2011/09/06  & 11283 &  N14 W18 & X2.1 & 22:12 & 22:20 & 22:23 & 11 & A & 102.9 & 990\tablefootmark{b}\\
        2012/03/07  & 11429 &  N18 E31 & X5.4 & 00:02 & 00:24 & 00:40 & 38 & B & 117.8 & 2680 \\
        2012/06/14  & 11504 &  S17 E06 & M1.9 & 12:51 & 14:35 &  15:56 & 185 & A & 117.4 & 990\\
        2019/03/08  & 12734 &  N11 W04 & C1.3 & 03:07 & 03:18 & 03:58  & 51 & A & 98.2 & 290 \\
        \noalign{\smallskip}
        \hline
\end{tabular}
\tablefoot{
The table contains data regarding NOAA Active Region and flare location, flare class, GOES flare time interval (from start to end, including the peak time), STEREO satellite used for the analysis, its separation angle with SDO, CME speed as listed in the LASCO catalog\\
\tablefoottext{a}{CME speed as listed in the LASCO catalog (\url{https://cdaw.gsfc.nasa.gov/CME_list/}).}
\tablefoottext{b}{The CME speed of the September 6, 2011, event is from \cite{dissauer2016projection} and measured from STEREO data.}
}
\end{table*}
Table \ref{tab: table-review} provides an overview of the first flare/CME event in each dataset, in particular by showing the time interval of the GOES flare by indicating start, peak, and end time, the flare location and class, and the CME speed. The type of STEREO satellite used for the observation and its angular separation from SDO are also indicated. Three events in the table (September 6, 2011, June 14, 2012, and March 8, 2019) are better observed by STEREO-A, while images from STEREO-B are used to observe the March 7, 2012, event from above the limb. 

The existing dimming studies on these events mainly concentrated on the impulsive and main dimming phase related to the CME's early evolution phase. Here we focus on the ``long-term'' evolution of dimmings, with the main aim of studying their recovery phase.  
The selected events vary in terms of flare class, CME speed, location on the Sun, and activity cycle. 
All events occurred on disk for SDO. 
The choice of the 3-day analysis time interval is made to better study the lifetime of the dimmings and to investigate whether the presence of subsequent eruptions from the same AR affects the dimmings end-of-life phase.
One important issue regarding the long-term evolution of dimmings (over several days) is the effect of the solar differential rotation and the overall changes of the corona during these times. 
Indeed, some distortions may be observed when applying derotation correction to compensate for the effect of the differential rotation.

\section{Analysis}
\label{Analysis}
To study its evolution in time, the dimming region is firstly identified following the threshold-based segmentation procedure described in \cite{dissauer2018detection}. For each case under study, a pre-event image is selected 30 min before the starting time of the first flare/CME in the dataset and used as a reference (base map). 
Next, all the maps in the dataset are rotated back in time to the base map to correct for differential rotation. 
However, the process of differential rotating each image back to the base map brings some distortion of the image itself in proximity to the solar limb. 
For what regards the eastern limb, the presence of artifacts is observed for long rotation time intervals. For the western limb, some part of the solar disk (depending on the time range of the differential rotation correction) is left as a blank segment (as there is no information from the far side of the Sun available). 
An example is the case of the September 6, 2011, event, where the AR at the beginning of the analysis is located close to the Sun's center but already in the western half of the disk. After 48 hours, part of the AR and the dimming has rotated onto the far side of the Sun. Therefore, for this event, we restricted the analysis to 48 hours, whereas in the other events it is done for 72 hours.

After the differential rotation is completed, base-difference (BD) images are created, to visualize absolute intensity changes, as well as logarithmic base-ratio (LBR) images, to visualize relative changes in intensity. BD images are calculated as the pixel-by-pixel difference between the image intensity at the generic time $t_n$ and at the reference time $t_0$, and LBR images as the logarithmic ($\log_{10}$) ratio of the intensities \citep{dissauer2018detection}. 
In this study, LBR images are used to identify the dimming region, while BD data are useful to extract relevant properties of the dimming region, namely area and brightness.
Finally, the images are cropped to a smaller field of view to create a sub-map containing only the AR of interest and the full extent of the dimming region.

\subsection{Dimming detection procedure}
\label{procedure}

Following the procedure described in \cite{dissauer2018detection}, a pixel is flagged as dimming pixel when its logarithmic ratio intensity drops below a certain threshold; that is, $-0.19$ DN.
However, because of noise in the images, especially during long time series as investigated here, not all the pixels below the threshold should be considered as dimming pixels. For this reason, segmentation of the image is used to smooth the region of interest and avoid noise. Among the segmentation techniques, region growing is used for medical and scientific applications; in particular, we implemented a region growing algorithm by using the IDL routine region\_grow.pro as a guideline. 
The algorithm identifies the dimming region starting from specific pixels, called seed pixels, defined as 10\% of the darkest pixels in each LBR image. The algorithm finds the 
four connected neighbors to each seed pixel and adds them to the dimming regions as long as they fulfill a certain criterion; that is, they have an intensity lower than the threshold value of $-0.19$ DN. Eventually, the final region grows by including all the connected neighbors and a binary dimming map is obtained. A binary dimming map is composed of a binary image where dimming pixels have a value of 255 and non-dimming pixels' value is 0, and the same metadata of the corresponding LBR map.
We decided not to use morphological operators to fill possible holes left by the algorithm on the images, since those holes may indeed be physically there in between different regions of the dimming.

\begin{figure*}[!h]
\centering
\resizebox{0.95\hsize}{!}{\includegraphics[scale=0.25,trim=200 100 0 150, clip]{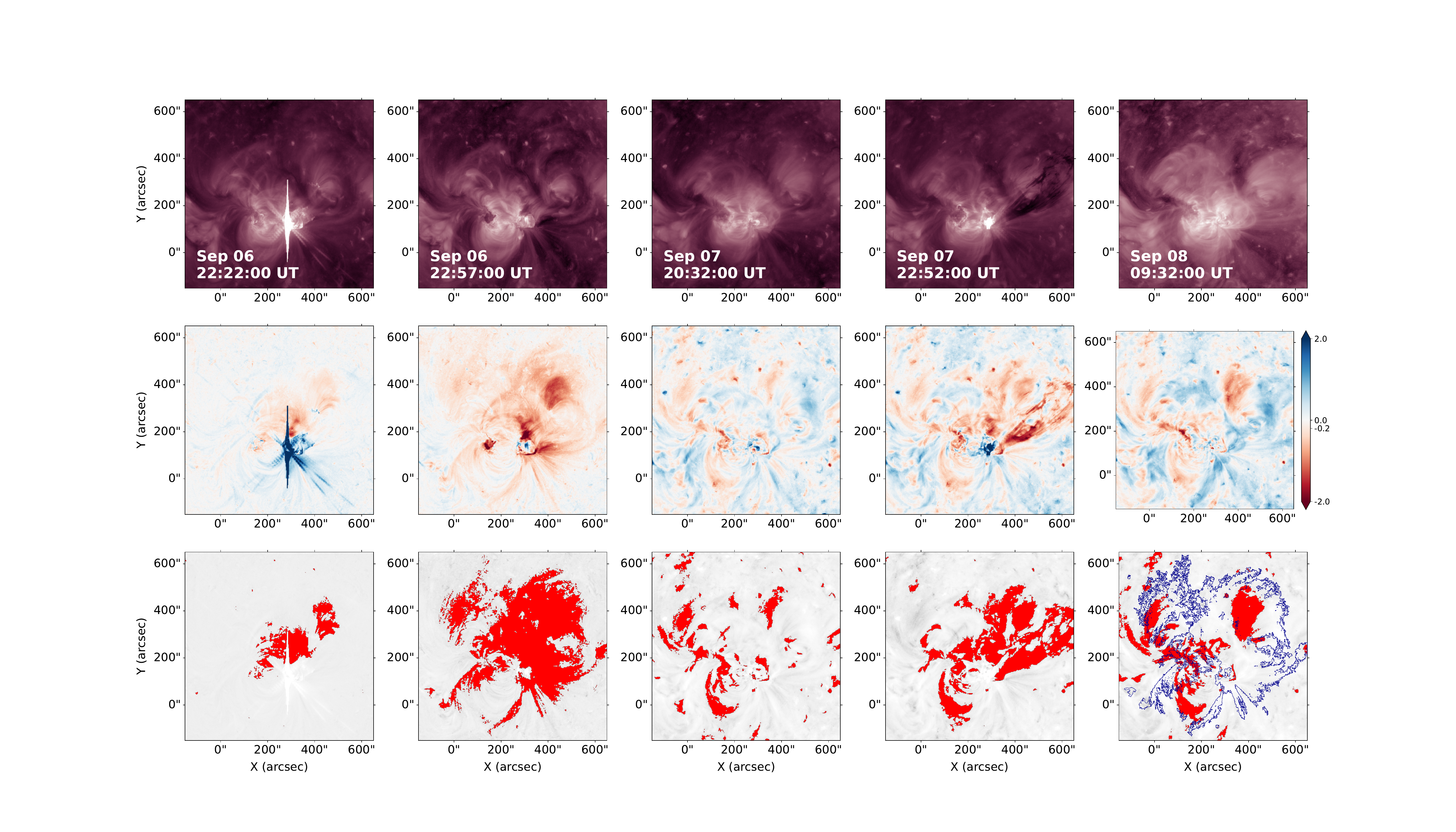}}
\caption{Snapshots of the evolution of the 
  coronal dimming region of the event on September 6, 2011. Top row: time sequence of SDO/AIA 211 \AA\ filtergrams. 
  Center row: corresponding Logarithmic Base-Ratio (LBR) images scaled to the range $-2$ to $2$ DN, highlighting both decreases (red) and increases (blue) in emission.
 Bottom row: instantaneous dimming pixel mask (in red) on top of the LBR image in grayscale. In the bottom right panel, the blue contour of the largest dimming mask within the first two hours of the event evolution is shown on top of the instantaneous dimming mask. The associated movie is available online.
  }
  \label{shot-sept}
\end{figure*}

\begin{figure*}[!h]
\centering
\resizebox{0.95\hsize}{!}{\includegraphics[scale=0.25,trim=200 100 0 150, clip]{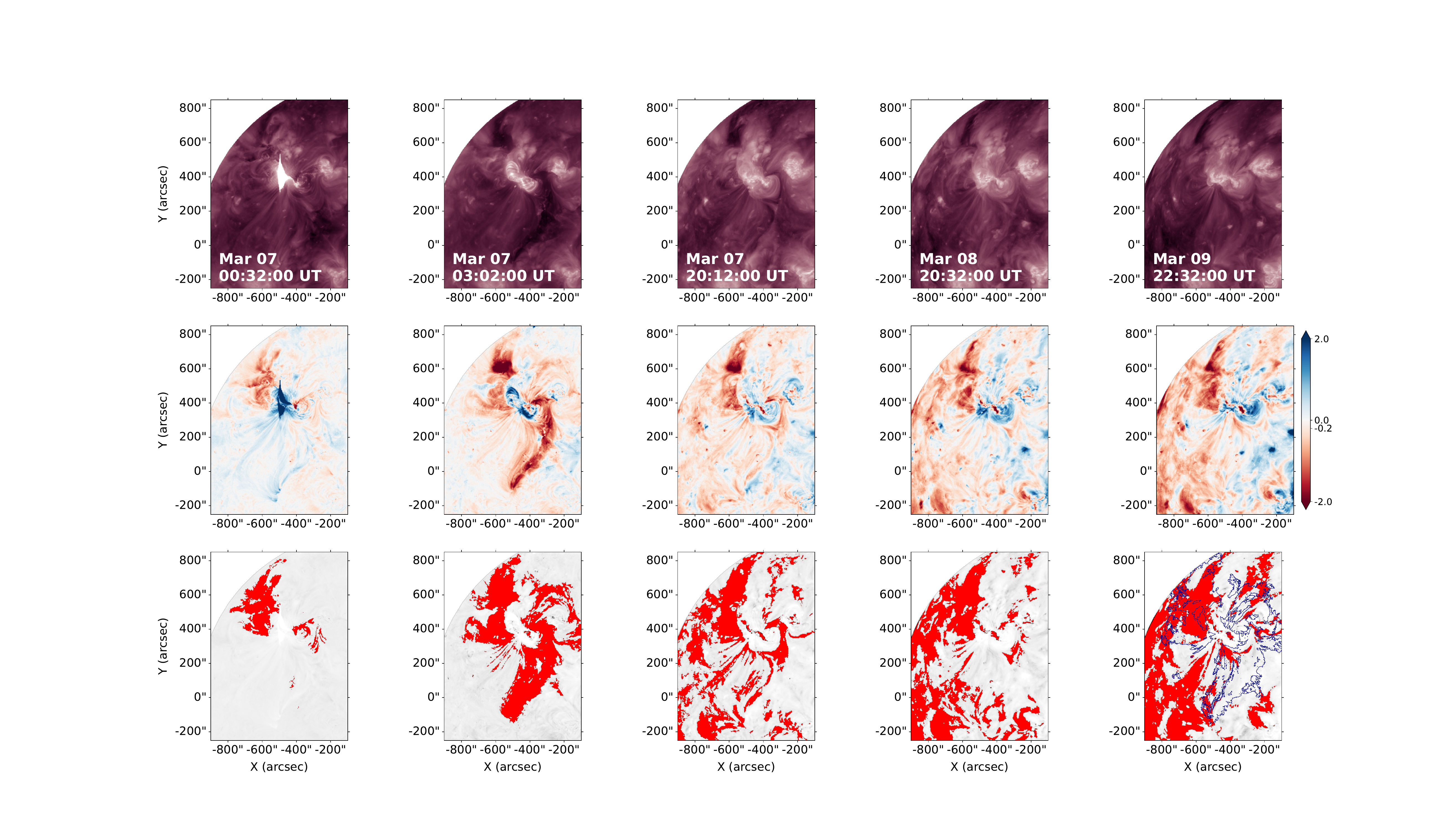}}
\caption{Same as in Fig. \ref{shot-sept}, but for the event on March 7, 2012. The time series covers 72 hours starting 30 min before the associated event, with a temporal cadence for the images acquisition of 5 min for the first 24 h and 30 min for the successive 48 h of the analysis. The associated movie is available online.}
\label{shot-march}
\end{figure*}

\begin{figure*}[!h]
\centering
\resizebox{0.95\hsize}{!}{\includegraphics[scale=0.25,trim=200 100 0 150, clip]{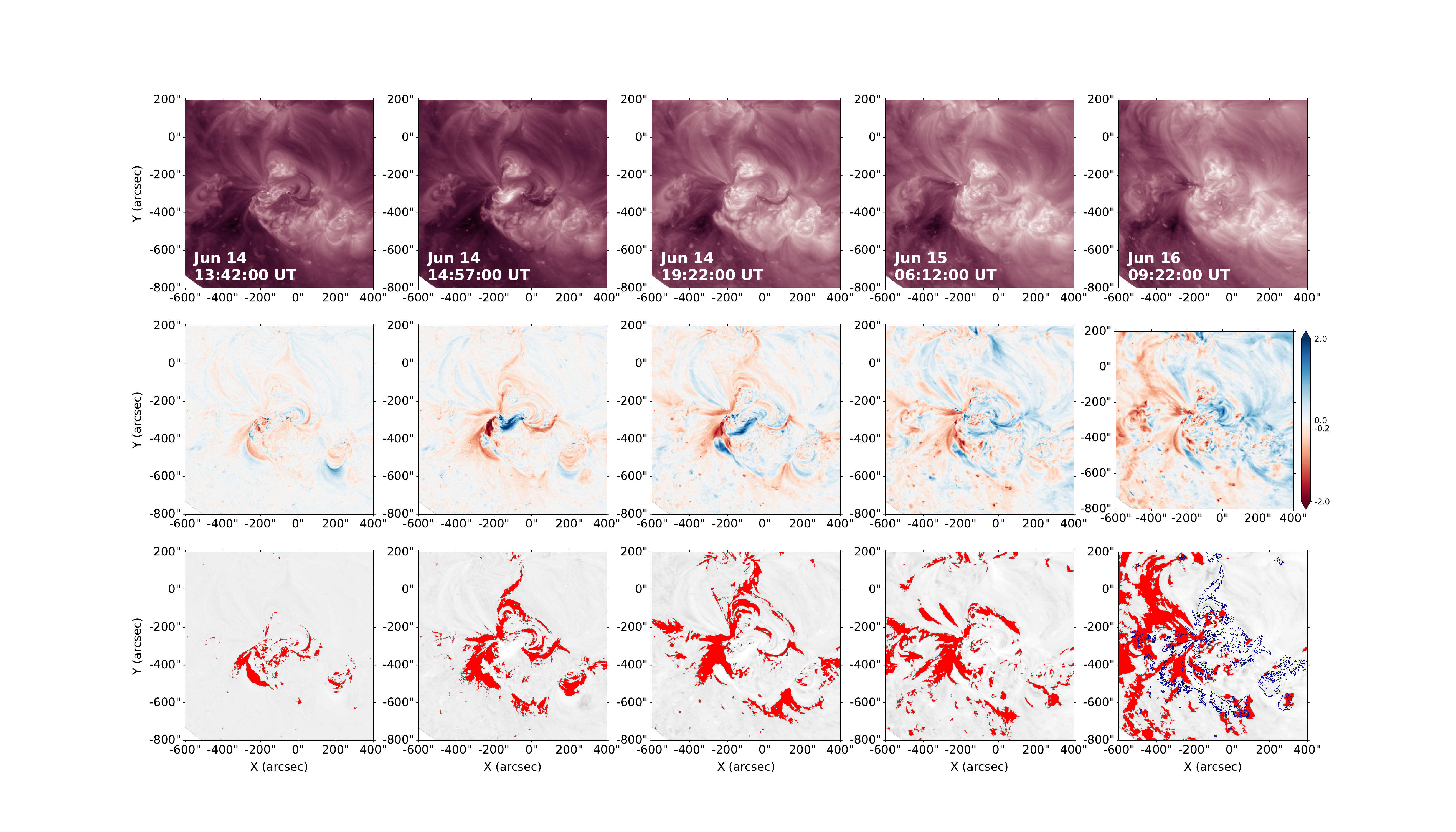}}
  \caption{Same as in Fig. \ref{shot-march}, but for the event on June 14, 2012. The associated movie is available online.
  }
\label{shot-june}
\end{figure*}

\begin{figure*}[!h]
\centering
\resizebox{0.95\hsize}{!}{\includegraphics[scale=0.25,trim=200 100 0 150, clip]{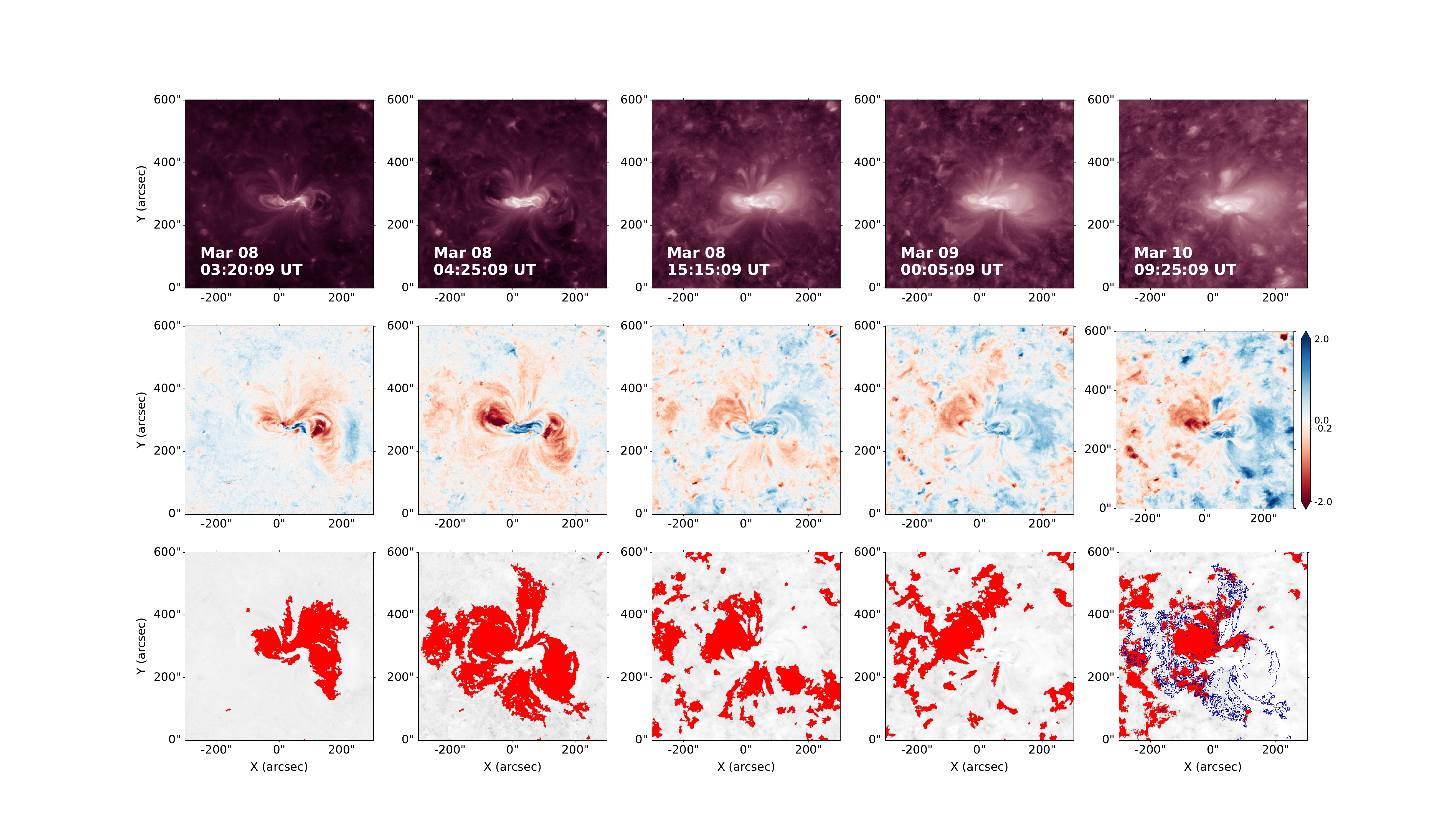}}
\caption{Same as in Fig. \ref{shot-march}, but for the event on March 8, 2019. The associated movie is available online.}
\label{shot-WD}
\end{figure*}

Figures~\ref{shot-sept} -- \ref{shot-WD} and the accompanying videos illustrate the time evolution of the four events under study.
In each figure, the top row shows AIA 211 \AA\ images at five different times during the analysis time interval while the center row shows the corresponding LBR images. The color bar marks the LBR intensity within the range of $-2$ to $2$ DN, indicating the intensity change relative to the base map: red indicates intensity decreases and therefore dimming regions, blue indicates intensity increases, the strongest ones are related to the flare in the form of flare ribbons and post-flare arcades, white indicates regions that do not change intensity compared to the pre-event image. In the bottom right panel, the contour of the largest dimming mask within the first two hours of the event evolution is shown on top of the instantaneous dimming mask to allow for a comparison between the initial eruptive phase and the end of the analyzed time range.
It is possible to already notice from these snapshots that the effect of de-rotation takes place for long observation intervals, the clearest example being Figure~\ref{shot-march} where red regions appear toward the eastern limb.

\begin{figure*}
  \centering
\includegraphics[width=0.8\textwidth]{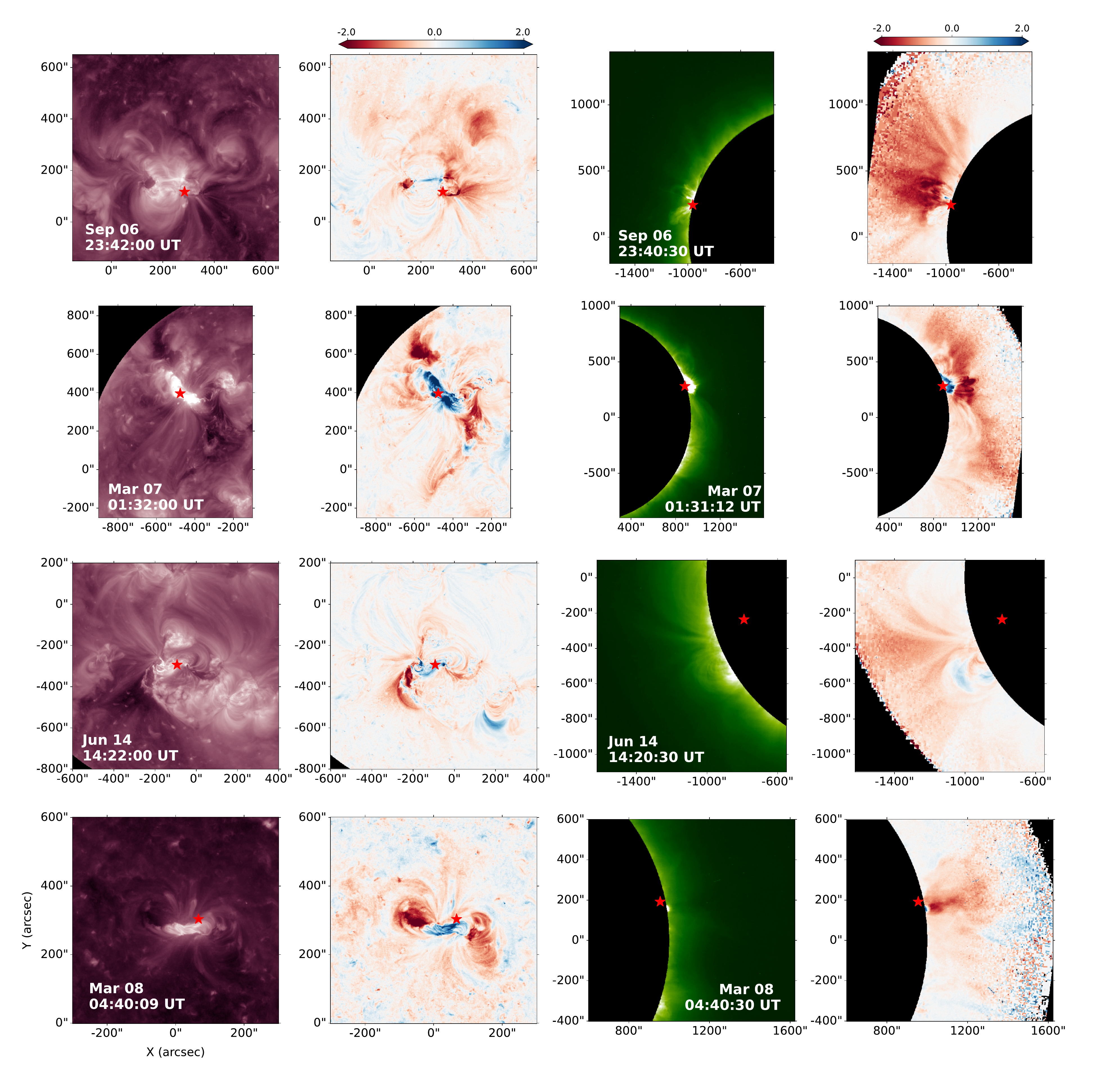}
  \caption{Snapshot of all the four cases under study. From top to bottom: September 6, 2011; March 7, 2012; June 14, 2012; and March 8, 2019. The shown time instant corresponds to when the dimming is close to its maximum extension. Left column: original SDO/AIA image. Middle left column: LBR image corresponding to the SDO/AIA direct image. Middle right column: original STEREO/EUVI image of the Sun. Right column: LBR image corresponding to the original STEREO/EUVI. The colorbar of the LBR images is based on the interval from -2 to 2, highlighting both decreases (red) and increases (blue) in emission. Red asterisks indicate flare coordinates for every event.
  }
    \label{sdo-stereo}%
    \end{figure*}

Figure~\ref{sdo-stereo} shows a snapshot of all the four cases under study, at a time instant when the main eruption dimming is close to its maximum extent. The column on the left shows the original SDO/AIA images, followed by the corresponding LBR image at the same time instant. The last two columns toward the right show the original STEREO/EUVI image of the Sun and the corresponding LBR image, respectively. The original SDO/AIA and STEREO/EUVI images have been calibrated and de-rotated to the base map time. 
The coloring scale of the LBR images is based on the same interval, from $-2$ to $2$ DN, to allow good visualization of the dimming. 

\subsection{Calculation of coronal dimming parameters}
The study of the four cases is based on the analysis of characteristic parameters, like brightness and area. 
In particular, we analyze quantities that describe the instantaneous state of the dimming (e.g., area and brightness), which are computed from the instantaneous dimming regions identified at each time step $t_i$. In addition, two new approaches are introduced regarding the brightness analysis. They make use of instantaneous brightness, but focus on specific areas of the image, to study the long-term evolution of the coronal dimmings. 
This is a more challenging task than for the impulsive phase of the dimming, as over the timescale of days the corona itself is changing, and the differential rotation correction of individual coronal pixels gets more uncertain, which is then reflected as noise in the BD and LBR images.

\subsubsection{Brightness}
\label{chap:br_analysis}

The binary dimming maps (cf. Sect. \ref{procedure}) are applied as a mask on top of the corresponding BD images so that only values corresponding to dimming pixels will be available. 
Then, it is possible to compute the instantaneous brightness of the dimming region as the sum of the intensity of non-masked pixels in the BD image.

Instantaneous brightness curves are a standard approach used to visualize the behavior of the dimming over time. However, this method is generally used for the analysis of shorter time periods after the solar flare/CME occurred, whereas the present analysis covers 3 days of observation. 
Such a long time of observation implies a widespread presence of artifacts related to the differential rotation process, which may be identified as coronal dimmings by the region-growing algorithm (most severe in regions close to the solar limb). 
Indeed, the detection of seed pixels at every instant may result in the selection of pixels related either to the artifacts or to an increasing dimming region generated after the eruption of other flares/CMEs that occur during the studied time interval. 
This leads to a continuous variation in the dimming area over time, which is no longer connected to the recovery process of the first eruption we focus on.
In order not to be affected by the dimming variation in size and to limit the influence of artifacts, we decided to focus the analysis on specific localized regions of the cropped image and hereby suggest two approaches to the analysis of the dimming lifetime.

The first method is the application of the same dimming mask to each map. 
Compared to a cumulative mask, an instantaneous fixed mask allows to minimize the number of flaring pixels that are present inside the mask and that may affect the calculations.
This ``fixed'' mask is chosen as the largest instantaneous dimming mask $D$, measured in a number of pixels $p_k$, among the LBR maps within the first two hours from the beginning of the flare/CME activity. 
This amount of time $t_{max}$ is considered enough to include the impulsive phase of the dimming \citep[cf., the statistics in][]{dissauer2018statistics, chikunova2020coronal}. 
However, in the cases of the June 2012 and March 2012 events, the chosen map is a relative maximum and not the absolute maximum extension. 
In both cases, the dimming area continues growing and, in particular, during the June 2012 case, the flare activity has a long duration, of approximately 3 hours. 
After identifying the maximum dimming extension within the impulsive phase and using it as a mask on top of BD maps, the instantaneous brightness $I_{f.m.}$ of each map (at time $t_m$) is computed as the sum of all the pixels' intensities inside the masked region:

$$I_{f.m.}(t_m) = \sum I(p_i, t_m )D_{fixed} (p_k , t_{max})$$
\\
where $I(p_i, t_m)$ is the intensity at $t_m$ for a certain dimming pixel $p_i$ within the fixed dimming mask $D_{fixed} (p_k , t_{max})$.
In this case, it is not considered whether a pixel is an instantaneous dimming pixel at the time $t_m$, but all the pixels of the fixed dimming mask contribute to the calculation.

The second approach focuses on investigating a very localized region of the overall map: several 3x3 pixel boxes within the dimming fixed mask are considered using the original SDO/AIA images, the BD images, and the LBR images, then the mean value of the intensity of the pixels inside each box is computed. 
The location of each box is manually selected by examining the LBR images, where dimming regions are easily identified by the eye, and at the dimming binary maps. Indeed, boxes can be placed close to the eruption center or further away and the evolution of brightness over time can be monitored. This approach is very versatile and its main advantage is that, depending on the location of each box, different behaviors can be observed. 
In addition, it allows to study the localized behavior in dimming regions which may be less ``smeared ou'' by the overall long-term evolution and data processing.

\subsubsection{Area}
\label{chap:area_analysis}
Regarding the area estimation, the algorithm developed by \cite{2023Chikunova} is used: it computes the extension of the solar surface by taking into account the 3D nature of the Sun since AIA images are projections of the solar sphere onto 2D planes. Thus, the area $A$ of the solar surface, having a projection on the circle delimited by the solar limb, is calculated by using a double integral over the surface:

$$A= \iint{\sqrt{1+ \left(\frac{dz}{dx}\right)^2 + \left(\frac{dz}{dy}\right)^2} dx dy}$$
\\
knowing the relation between the coordinate set ($x, y, z$) and the solar radius $R^2 = x^2 + y^2 + z^2$ . The information about the exact radius and the ratios to convert from km to arcseconds on the Sun and to convert arcseconds to image pixels unit are stored in the metadata of the images. The integrals are solved and the output is a surface image of the Sun, 
where the value of each pixel belonging to the solar disk is the area extension of that specific portion of the solar sphere, in km$^2$. As was done in Sect. \ref{chap:br_analysis}, by applying the dimming maps as masks on top of the surface maps and by summing the values of pixels, we are able to retrieve the instantaneous dimming area for each time step (map).

\section{Results}
\label{Results}

In the following subsections, we give a detailed analysis of the global dimming evolution for each of the four events under study. 
We also present results for localized areas within the global dimming region, to highlight the fine structure within the dimming regions and to compare the long-term evolution of the global dimming properties and its localized behavior.
Finally, a comparison among all the events is given, by quantifying the global dimming intensity evolution with the fixed mask approach.

\subsection{September 6, 2011}
\label{chap:result_sept}

\begin{figure}[!ht]
  \centering
  \includegraphics[scale=0.5,trim=0 10 0 60, clip]{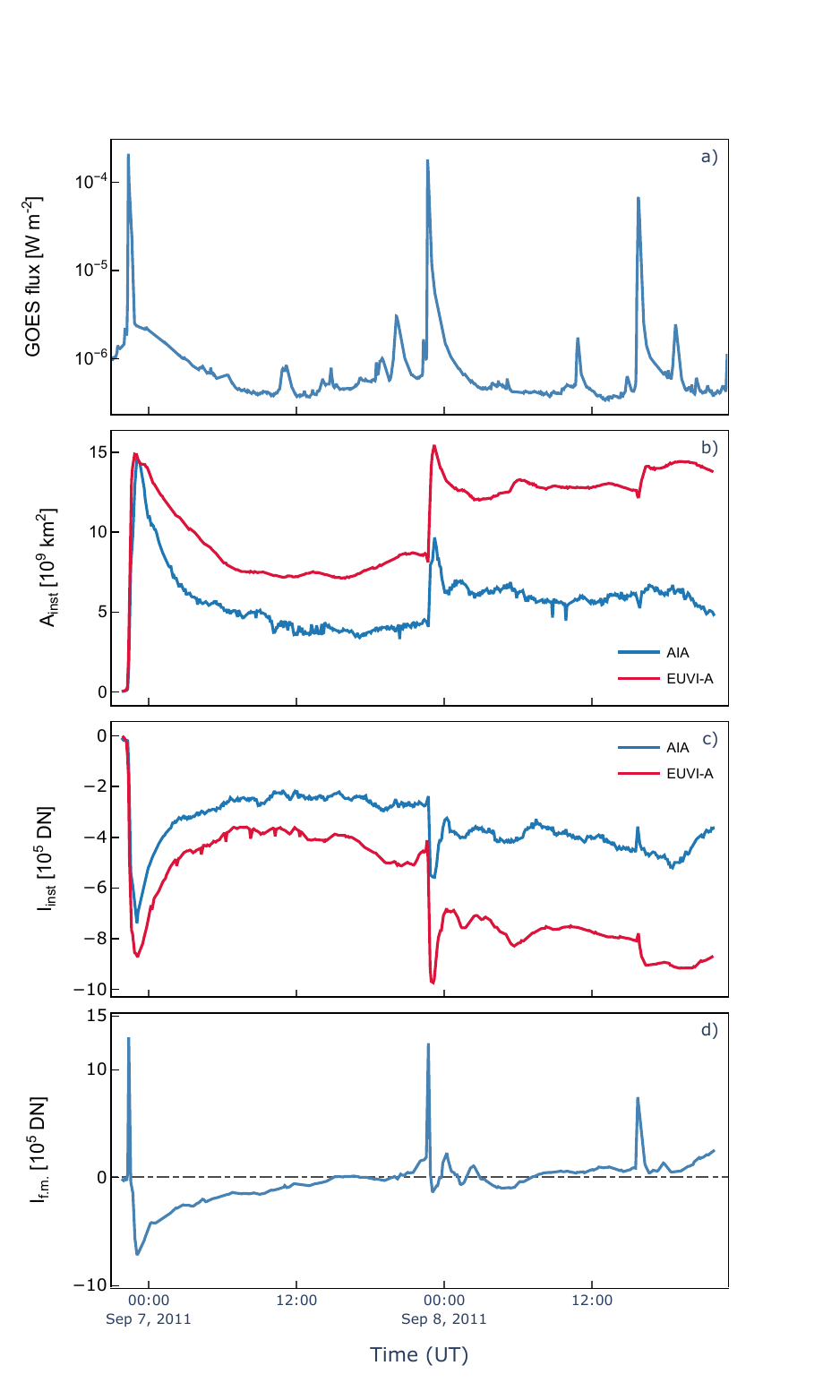}
  \caption{Evolution of dimming area and brightness in the September 6, 2011, event. The data from STEREO, having a higher cadence, are smoothed. (a) GOES 1-8 \AA~soft X-ray flux. (b) Instantaneous dimming area evolution over time for SDO/AIA 211\AA~data (blue curve) and STEREO/EUVI-A 195\AA~data (red curve). (c) Instantaneous dimming brightness evolution over time. Showing the comparison between SDO/AIA (blue curve) vs STEREO/EUVI-A (red curve). (d) Instantaneous brightness evolution within the fixed mask. The dashed horizontal line marks the pre-event brightness level.}
    \label{sept-goes}%
\end{figure}

Figure~\ref{sept-goes} shows the outcome of the analysis for the September 6, 2011, event. The four panels show the flare evolution using the GOES 1-8 \AA~soft X-ray flux (panel a), the instantaneous dimming area $A_{inst}$ (panel b), the instantaneous brightness of the dimming region $I_{inst}$ (panel c) and the total brightness of the region within the fixed mask $I_{f.m.}$ (panel d) as a function of time using SDO/AIA 211\AA~data. These intensities are related to the observation of the event in the base-difference images. 
Panels b and c also contain a comparison of instantaneous area and brightness between data from SDO/AIA and STEREO/EUVI in 195~\AA.

The GOES flux helps in visualizing the time of occurrence of the flares, which in general occurs close in time to the initiation of the associated CME \citep{marivcic2007acceleration, bein2012impulsive}. The analysis covers 48 hours and, during this time, three flares occur, each associated with a CME. The instantaneous brightness (Fig.~\ref{sept-goes}c) shows a recovery, especially after the first and second flare, but never reaches a full recovery to the pre-event level. The same holds for the instantaneous area (Fig.~\ref{sept-goes}b).
Data from EUVI, onboard STEREO-A, are employed as well and they show a matching trend with the ones from AIA. 
In particular, the steep decrease (increase) related to the dimming brightness (area) exhibits a similar behavior and the peak times are almost co-temporal. Also in the EUVI-A evolution, there is no full recovery of the dimming. 

Figure~\ref{sept-goes}d shows the behavior of the instantaneous brightness within the fixed dimming mask. 
In this case, both dimming and non-dimming pixels for the considered time step are contained in the identified region, hence bright pixels belonging to the flares contribute to the brightness computation. The three flares/CMEs appear like three spikes and after the first one a sharp decrease in brightness is observed, which falls significantly below the pre-flare level (corresponding to the so-called ``post-flare dimmings'' in full-Sun observations; see \cite{veronig2021detection}). The brightness decreases until a minimum point, corresponds to the maximum extension of the dimming region. 
During the hours past the first eruption, it is possible to observe a complete recovery of the brightness on September 7 at 14:57 UT, far before 24 hours passed. 
Even when the second flare on September 7 occurs (24 hours after the first one), it does not substantially affect the intensity within the fixed dimming region identified from the first event. 
This is related to the dimming location and direction of its expansion: the first dimming expanded mostly toward the northwest of the eruption site, while the second one expanded both in that direction and toward the northeast, thus being partially located outside of the fixed dimming mask. 
The part of the second dimming that still remains inside the fixed mask is smaller compared to the first dimming extension, so the total brightness of the fixed region oscillates around the zero value, with both dark dimming pixels and brighter pixels present within the fixed mask region. Lastly, a smaller spike corresponding to the last flare is observable on September 8. The detected dimming is very small in this case and overlaps with parts of the previous dimming regions that have not disappeared yet. 

By carefully examining SDO/AIA 211\AA~snapshots and videos, we observe that the dimming gradually dissolves due to bright coronal loops slowly expanding in the AR (blue in LBR images).
In particular, the dimming region located furthest from the eruption center is the first to recover, followed by the nearby northwestern parts, while some dark dimming regions still remain close to the center of the eruption. This is further investigated by analyzing the recovery behavior of certain small regions by applying the pixel boxes approach. 

\begin{figure*}[h]
\centering
  \resizebox{0.9\hsize}{!}{\includegraphics[scale=0.85]{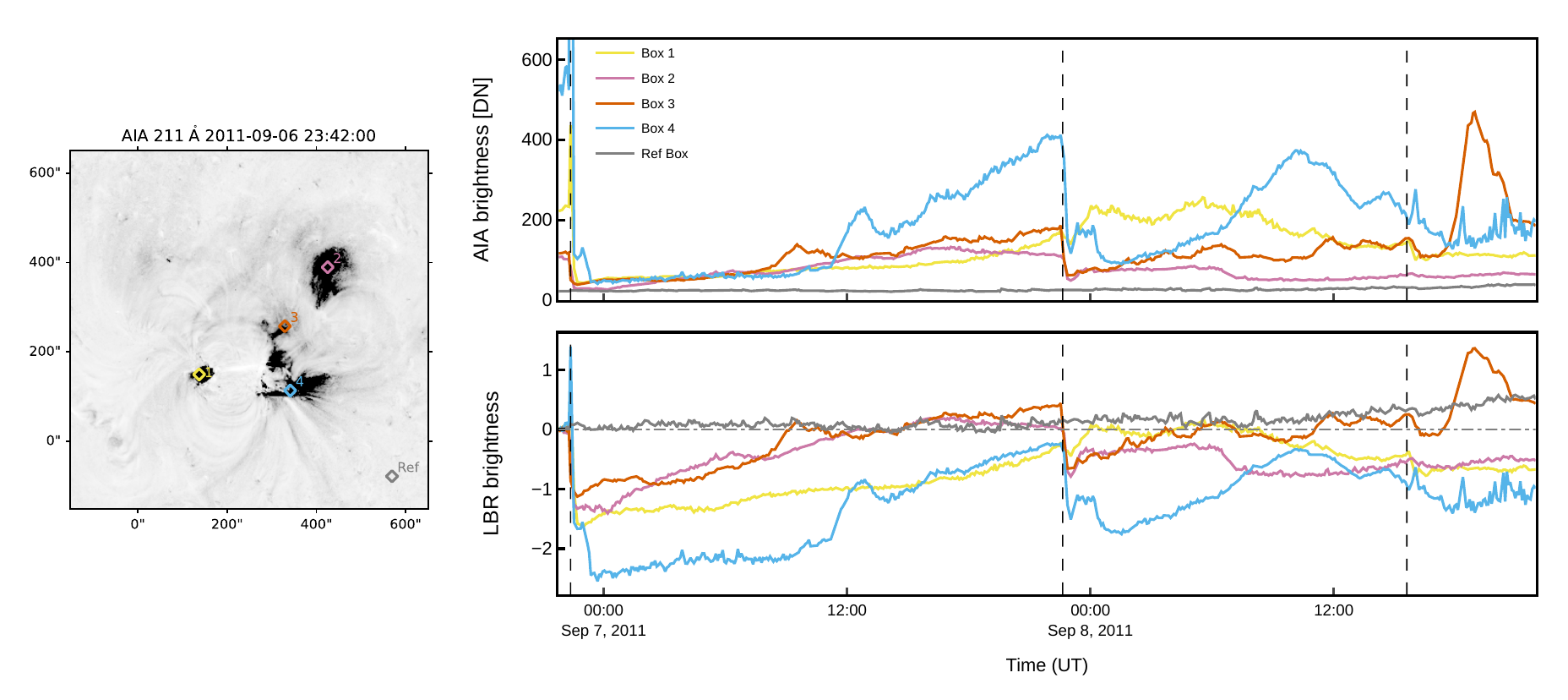}}
  \caption{Locations of the chosen area boxes and comparison of their brightness behavior for September 6, 2011, event. Left panel: SDO/AIA 211 \AA~logarithmic base ratio images during the impulsive phase of the dimming. Four colored boxes of size 3x3 pixels in different dimming regions of interest are over-plotted. The reference box is located in a random quiet Sun region. Right panel: Intensity evolution of the boxes from the original AIA images (top) and LBR images (bottom). The dashed horizontal line marks the pre-event brightness level. The dashed vertical lines indicate the peak time of the two X-class and the one M-class flare.}
  \label{im:plot box-sept}%
\end{figure*}

The left panel of Figure~\ref{im:plot box-sept} shows the LBR map on September 6, 2011 during the impulsive dimming phase, with four colored diamonds that mark the center of the 3x3 pixel boxes. In particular, boxes 1 (yellow), 3 (red), and 4 (light blue) are positioned in proximity to the eruption site, while box 2 (pink) lies further away, in the middle of a very dark dimming region, most likely a secondary dimming region within a magnetic plage region. 
The goal of the boxes is to keep track of the ongoing dynamics and intensity of behavior in the chosen areas. 
The right panel shows the evolution of the average original and logarithmic base-ratio brightness of the four boxes. 

\begin{figure}[ht]
  \centering
  \includegraphics[scale=0.6]{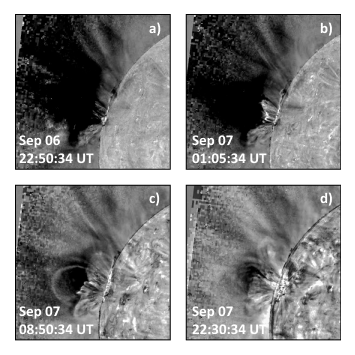}
  \caption{
  Snapshots of coronal loop expansion in STEREO-A/EUVI 195 \AA\ base difference images for the event on September 6, 2011. The time steps are: (a) maximum dimming extent, (b) appearance of post-flare arcade and faint loops, (c) further expansion of loops into the higher corona, and (d) just before the second flare. The full temporal evolution is available in the accompanying video created in JHelioviewer \citep[][]{Mueller:2017}.
  }
  \label{im:snapshots}
\end{figure}

It is seen that box 2 and box 3 already show recovery to the pre-event level before the occurrence of the second flare (marked by the dashed line). 
The corresponding AIA movies show a gradual expansion of post-flare and peripheral loops into or toward the dimming region of box 3 and later on into box 2.
For this event, the STEREO-A satellite was in a good position to clearly observe the expansion of the coronal loops from the side. Figure~\ref{im:snapshots} shows snapshots of the STEREO/EUVI 195 \AA~base-difference movie illustrating the progression from the period of the maximum extent of both off-limb and on-disk dimmings to their near disappearance just before the onset of the second flare: (a) the moment of maximum dimming extent, (b) the appearance of a post-flare arcade in the northern part of the dimming and faint loops in the southern part, (c) the expansion of loops into the higher corona, and (d) coronal loops are covering the off-limb dimming region before the second flare. The video further reveals that at the onset of the second flare/CME, the coronal loops get disrupted, dimmings reappear due to the evacuation of mass, and then the loops cover the dimming again, repeating the process.

Box 1 shows a different evolution. At the beginning, it shows a similar recovery rate as box 3 and we also observe post-flare loops expanding in this region, from the opposite site. However, the intensity within this box stays still reduced after that. 
The dimming region, where the box is located, shrinks by the expansion of flare ribbons into it, which could be indicative of reconnection between the overlying field and the erupting flux rope. 
Lastly, box 4 shows a weaker recovery in intensity during the first 12 hours after the CME, especially when compared to box 2. This result is in agreement with \cite{attrill2008recovery} and hence it makes sense to classify this dimming area as a core dimming region, from an observational point of view. 
However, it should be noted that this region does not correspond to the classical definition of core dimming; that is, footpoints of an erupting flux rope. This box is localized within a dimming region that is hypothesized to contain field lines that opened up during the eruption \citep{prasad2020magnetohydrodynamic}. 
Combining observations and data-driven MHD simulations, \cite{prasad2020magnetohydrodynamic} found that this dimming region is co-spatial with the footpoints of the dome surface of a 3D magnetic null point participating in the main flare/CME event. A causal connection between the magnetic reconnections at the 3D null and the dimming, due to the transformation of field lines of the inner spine to open field lines of the outer spine, was concluded in this study.

\subsection{March 7, 2012}
\label{chap:result_mar}

\begin{figure}[!ht]
\centering
\includegraphics[scale=0.5,trim=0 10 0 50, clip]{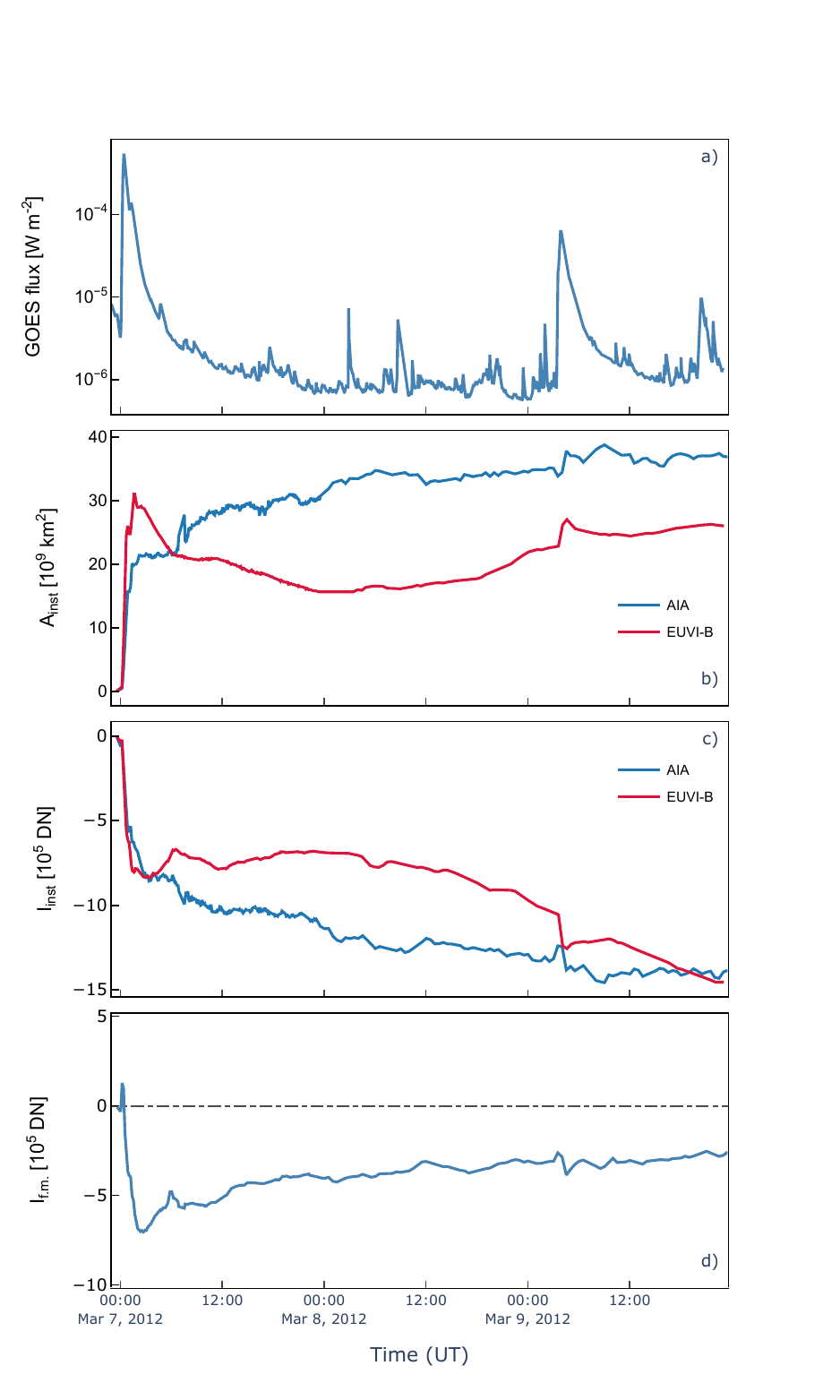}
\caption{Same as in Fig.~\ref{sept-goes}, but for the event on March 7, 2012. EUVI data are from STEREO-B.}
\label{fig: march-goes}%
\end{figure}

\begin{figure*}[!ht]
\centering
\resizebox{0.75\hsize}{!}{\includegraphics[scale=0.6, trim= 60 0 30 30, clip]{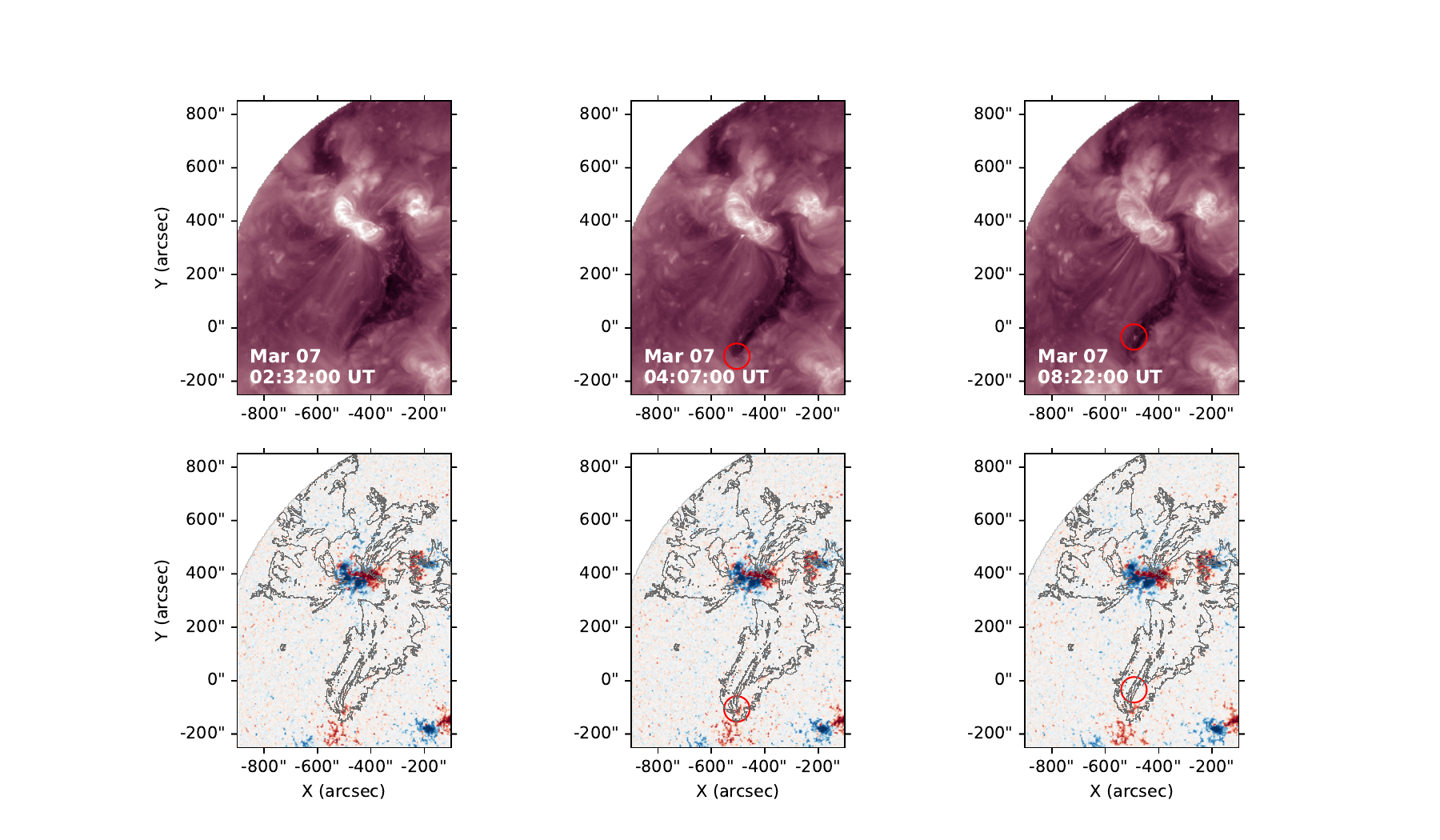}}
\caption{SDO/AIA 211\AA~original images (top panels) and HMI LOS magnetograms (bottom panels, scaled from $-750$ G to $+750$ G) on March 7, 2012, at 02:32, 04:07, and 08:22~UT. The red circles mark observed brightenings in AIA. The gray contours present the contour of the largest instantaneous dimming mask used for the analysis of the dimming intensity evolution. 
They indicate flux emergence and interchange reconnection inside the dimming region.}
\label{fig: march-bright}%
\end{figure*}

\begin{figure*}[h]
\centering
  \resizebox{0.9\hsize}{!}{\includegraphics[width=0.8\textwidth]{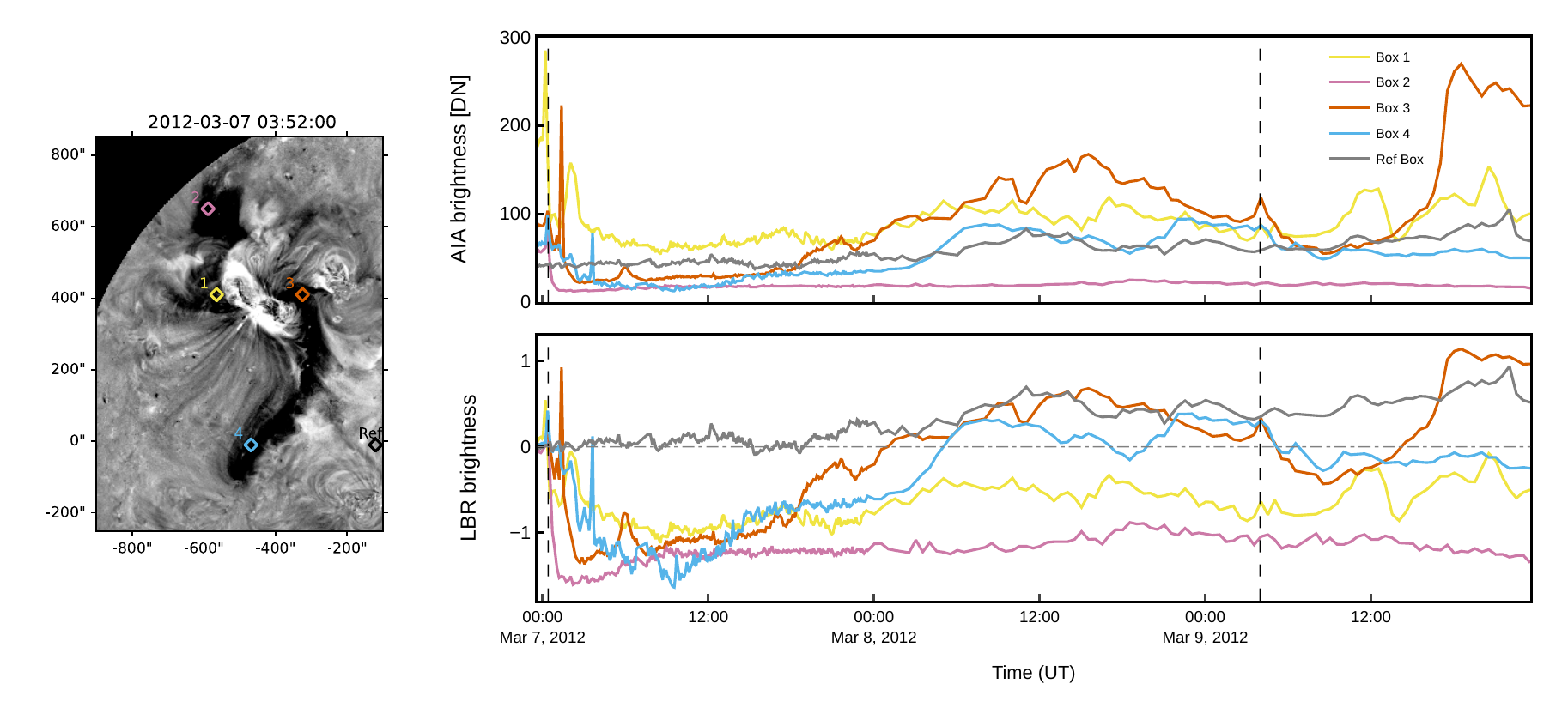}}
  \caption{
  Same as Fig.~\ref{im:plot box-sept}, but for the March 7, 2012, event.}
  \label{im:plot box-march}%
\end{figure*}

Fig.~\ref{fig: march-goes} shows the evolution of dimming brightness and area for the 2012 March 7 event. From the instantaneous dimming area and brightness curves in panels b and c, a continuous growth of the dimming area alongside a decrease in brightness, in both SDO/AIA and STEREO-B/EUVI data is observed.
It appears that the dimming that formed after the flare/CME on March 7 does not recover after two days but continuously grows. At the time of the flare/CME on March 9, this trend is still preserved. Data from both the AIA and EUVI instrument show similar behavior, especially in the sharp increase (decrease) in area (brightness) and in the overall decrease in the instantaneous brightness. 
However, there is a difference between the observations made by the two instruments: while data from SDO/AIA show continuous growth in size and decrease in brightness, STEREO-B/EUVI data show a shrinkage of the dimming area (red line in panel b), which is accompanied by a mild recovery trend of the brightness, then followed by an almost steady value of both instantaneous area and brightness. This behavior is maintained until the second flare occurs after around 52 hours, which results in a further decrease in instantaneous intensity and growth of the area. 
By checking the original images and the instantaneous dimming maps, it appears that the dimming expands close to the eastern limb, further increasing its size with time. 
Due to the proximity of the AR to the eastern limb, this behavior is interpreted as an artifact of the differential rotation procedure which affects the two regions of the dimming, the one in the northeast, closer to the limb, and one in the southwest.
Therefore, using the fixed mask curve for the interpretation of this event is more reliable. 

By looking at the evolution inside the fixed dimming mask (Fig.~\ref{fig: march-goes}-d), the brightness does not recover to the pre-event level. 
The second flare does not appear as a distinct peak in intensity, in contrast to what happens for the September case (Fig.~\ref{sept-goes}-d). 
This is partially due to the simultaneous presence of dark dimming pixels still being the majority inside the fixed mask, but also due to the bright flare region being outside the fixed mask, contrary to the September 6, 2011, event.

Another interesting aspect of the March 7, 2012, event is revealed when examining the original SDO/AIA images and the dimming evolution within the fixed mask. As was already mentioned, this dimming is composed of two regions. While the region in the northeast does not recover within 72 hours of observation, the region in the south does. In particular, it is observed that the recovery does not happen only because of the expansion of coronal loops, like in the September 6, 2011, case,
but there may be also some emerging flux from below the photosphere and interchange reconnection inside the dimming regions that helps to refill the area: brightenings are observed at specific times inside the southern part of the dimming; for example on March 7, 2012, at 04:07, 05:42, 08:22, or 09:17 UT. Figure~\ref{fig: march-bright} shows two out of the above-mentioned time instants where brightenings are visible. These brightenings are observable in the original AIA image, but are also not included in the dimming region by the region growing algorithm, meaning that they are brighter compared to the surrounding region (i.e., the dimming): some bright pixels actually appear underneath the dimming.

The left panel of Figure~\ref{im:plot box-march} shows the LBR map on March 7, 2012, during the impulsive dimming phase, with four colored diamonds that mark the center of the 3x3 pixel boxes. A reference box is positioned in a quiet coronal region for comparison. In particular, boxes 1 (yellow) and 3 (red) are positioned in proximity to the eruption site, while boxes 2 (pink) and 4 (light blue) lie further away, presumably within secondary dimming regions \citep[cf.][]{vanninathan2018plasma}. Since the goal of the boxes is to keep track of the ongoing dimming dynamics, the chosen areas are strategic to monitor the intensity behavior on the opposite side of the twin dimmings, as shown in the right panel of the same Figure.

Out of the four boxes, only boxes 3 and 4 show a recovery to the pre-event levels between 24 hours and 36 hours from the first flare time. In contrast, boxes 1 and 2 remain dark throughout the entire analysis time. However, the March 7, 2012, case, particularly the upper region of the dimming, is strongly subjected to artifacts induced by the differential rotation procedure because of its proximity to the eastern limb. Box 3 shows coronal loops close to the center of the active region gradually expanding into the region. Box 4 is located in proximity to the observed brightenings highlighted in Figure~\ref{fig: march-bright}. It slowly reaches its minimum and then recovers with a two-step behavior.

\subsection{June 14, 2012}
\label{result_jun}

\begin{figure}[!ht]
  \centering
  \includegraphics[scale=0.5,trim=0 10 0 50, clip]{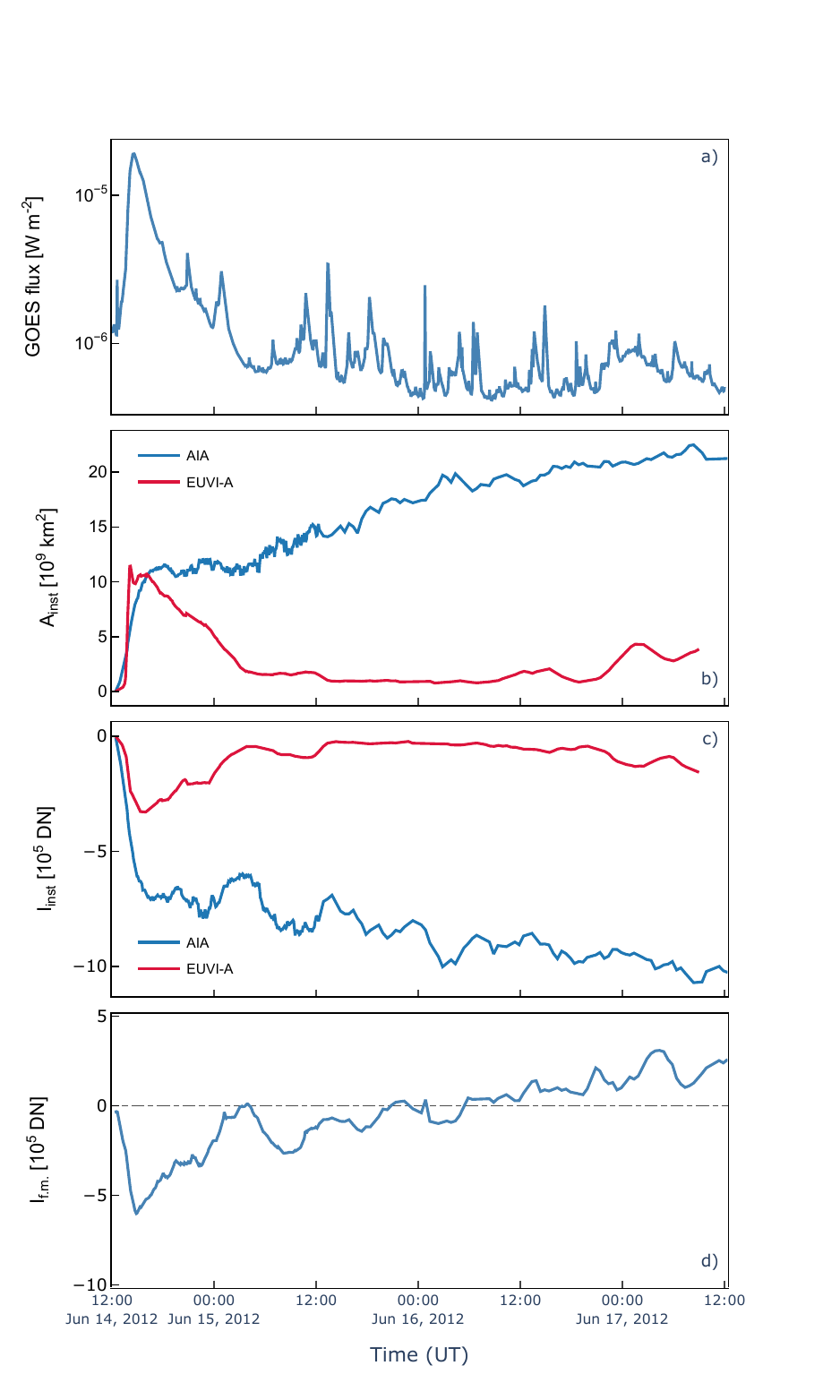}
  \caption{Same as in Fig. \ref{sept-goes}, but for event on June 14, 2012. 
  }
    \label{fig: june-goes}%
    \end{figure}

Results for the June 14, 2012, event are shown in Figure~\ref{fig: june-goes}. 
A huge discrepancy in the behavior of the curves obtained using the instantaneous dimming masks from SDO/AIA and STEREO-A/EUVI (panels b and c) is observed. A recovery back to the pre-event level is observed by the latter, while the former shows an incessant growth of the dimming. 
The June 14, 2012, event shows artifacts related to the differential rotation procedure resulting in the erroneous detection of dimming areas close to the solar limb. By changing the observational approach, in other words by limiting the observation to the region within the fixed mask (panel d), it is possible to see that the behavior of the corresponding intensity curve has a trend similar to the instantaneous brightness obtained with STEREO-A. A recovery back to the pre-event level is registered at around 03:40~UT, which is then followed by an oscillating behavior of the average brightness. 
By also looking at the original images, it can be concluded that there is some ongoing dynamics in the AR and its surroundings, which makes the interpretation of this event challenging. This complex dynamics could be related to ongoing reconnection activity, but further analysis, beyond the scope of this paper, would be required to prove that. The oscillating behavior of the brightness evolution within the fixed mask supports this interpretation, as dimming regions form throughout the full 72-hour time range.

\begin{figure*}[!ht]
\centering
\resizebox{0.9\hsize}{!}{\includegraphics[width=0.9\textwidth, clip, trim = 10mm 20mm 10mm 20mm]{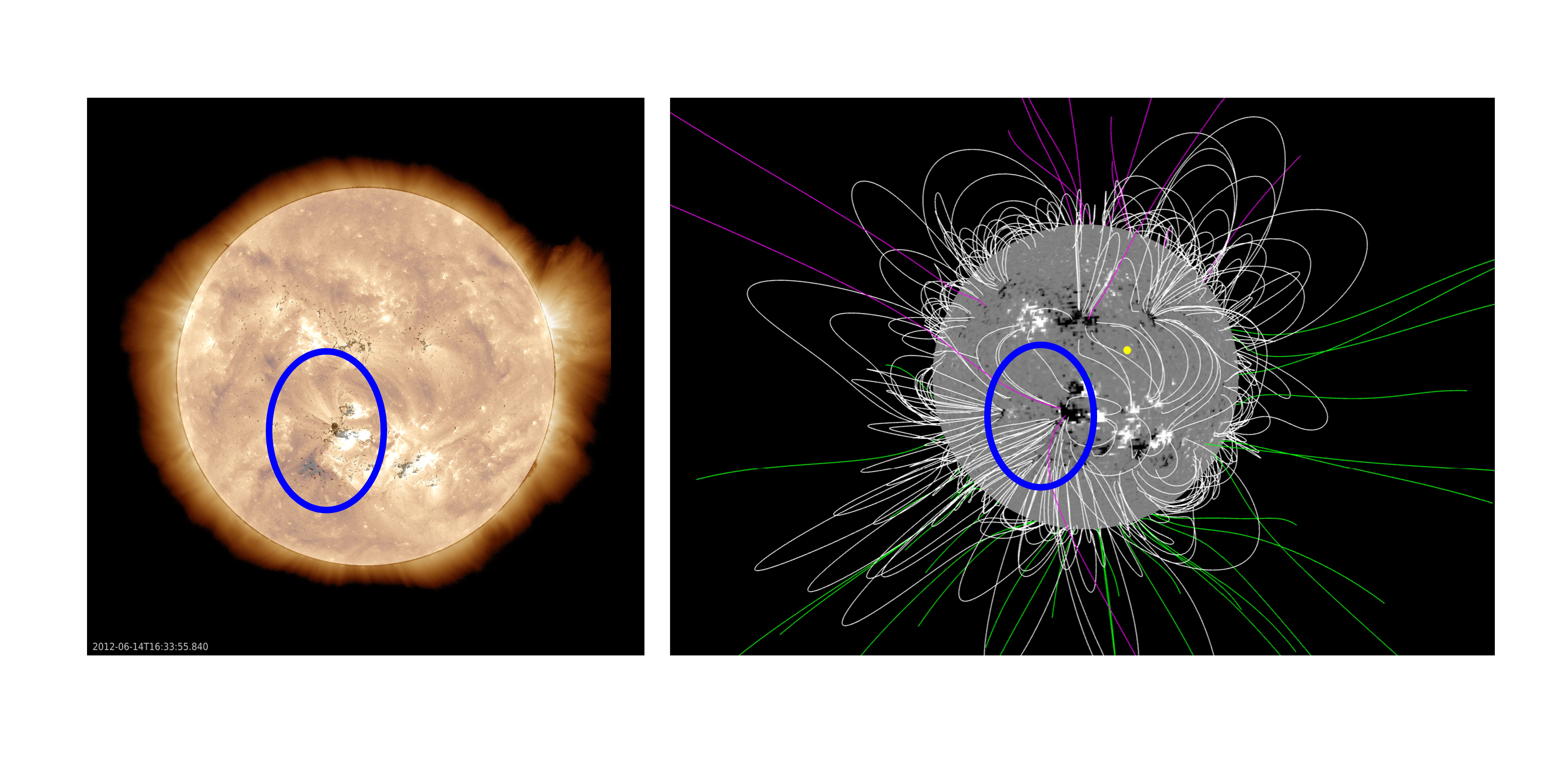}}
\caption{Magnetic field structure of AR 11504 and its surroundings for June 14, 2012, event. Left: Blend of snapshots of SDO/AIA 193~\AA~and the HMI line-of-sight magnetogram. The active region of interest and the suspected semi-open structure are highlighted by the blue circle. Right: Characteristic field lines of a PFSS extrapolation created with the \href{https://suntoday.lmsal.com}{Interactive Solar Fieldline Viewer}  by LMSAL are based on the models described in \cite{Schrijver:2003}. Open field lines (magenta lines) are present at the east of the active region of interest (marked with a blue circle), close to the location where we suspect a CH/open magnetic field structure, based on the dimming recovery behavior.}
\label{fig:pfss}%
\end{figure*}

In addition, AR 11504 is located close to a region of reduced emission, presumably a semi-open structure, which may evolve into a coronal hole (CH) in 2-3 days. 
Although, no coronal hole was identified with existing coronal hole detection algorithms--- tested with the coronal hole identification via multi-thermal emission recognition algorithm \citep[CHIMERA; ][]{garton2018automated}, the Spatial Possibilistic Clustering Algorithm \citep[SPOCA; ][]{verbeeck2014spoca}, and the Coronal Hole RecOgnition Neural Network Over multi-Spectral-data \citep[CHRONNOS; ][]{jarolim2021multi}--- a potential field source surface (PFSS) extrapolation shows open field lines at the location of reduced emission (see Figure~\ref{fig:pfss}). Jet-like features indicative of ongoing interchange reconnection are observed at the border of this region (see the accompanying video).

\begin{figure*}[h]
\centering
  \resizebox{0.9\hsize}{!}{\includegraphics[scale=0.8]{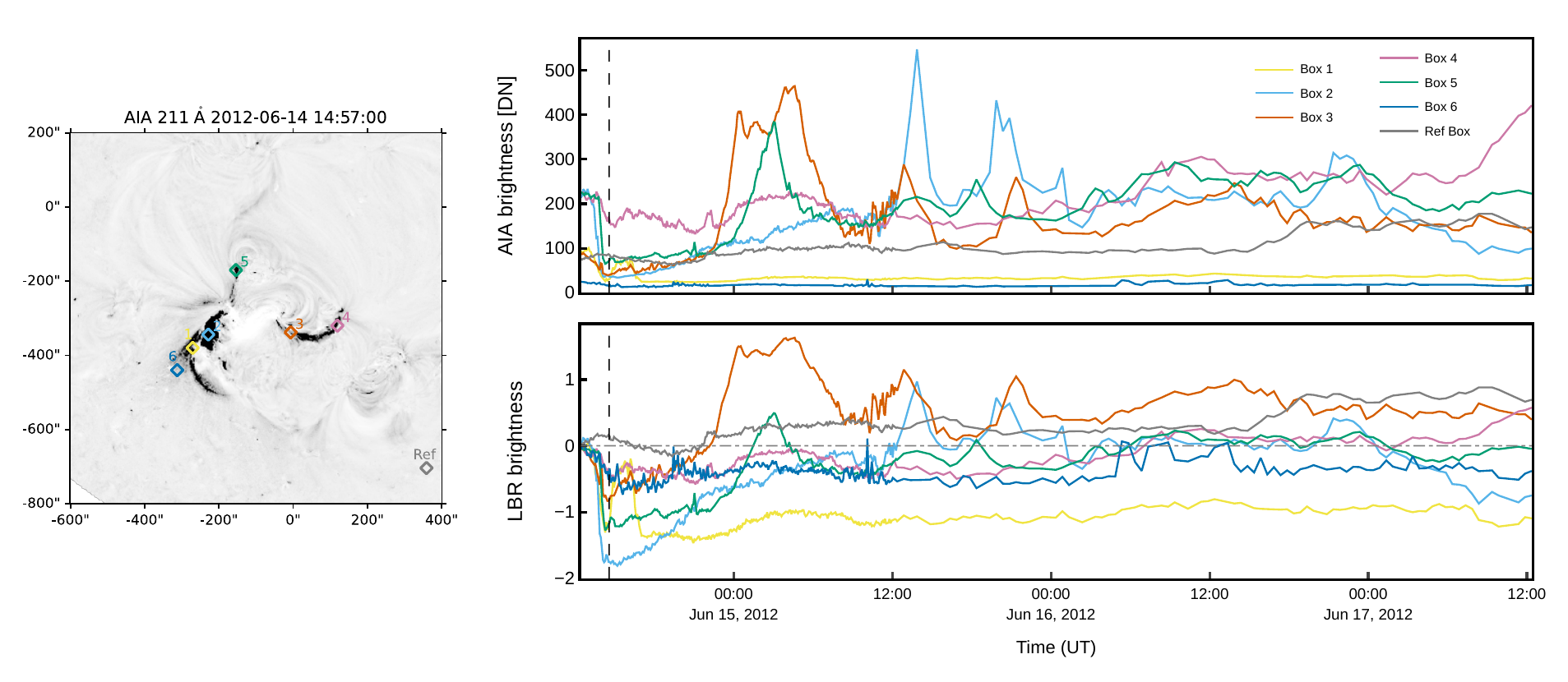}}
  \caption{
  Same as Fig.\ref{im:plot box-sept}, but for the June 14, 2012, event. Box 6 corresponds to a region of open magnetic field according to a PFSS extrapolation. 
  }
  \label{im:plot box-june}%
\end{figure*}

Applying the pixel box approach, we investigate the dimming recovery in different regions of interest. The left panel of Figure~\ref{im:plot box-june} shows the LBR map on June 14, 2012, during the impulsive dimming phase, with six colored diamonds that mark the center of the 3x3 pixel boxes. Boxes 1 (yellow) and 2 (light blue) are positioned inside the southeastern region of the coronal dimming, while boxes 3 (red) and 4 (pink) are situated to the west of the source, and box 5 (green) is located to the north of the source. Box 6 (blue) is placed inside the semi-open structure to the southeast of the AR. To validate the method and for comparison, a reference box is positioned in a random quiet Sun area.

The right panel of Figure~\ref{im:plot box-june} shows the evolution of the brightness inside the boxes. It can be observed that boxes 2-5 recover to the pre-event level within one day after the flare/CME event. In contrast, box 6 shows a gradual decrease in emission which reaches its minimum shortly after the flare peak, and stays at a reduced emission throughout the analysis time range. 
Box 1, which is closest to the semi-open region, shows the minimum relative intensity among all boxes, matching the intensity of box 6. Following a brief brightening caused by flare-related emission, box 1 consistently maintains this minimal intensity level throughout the whole observation period. 
Examining the movie of Figure~\ref{shot-june}, it shows bright loops that covered this region prior to the CME, entirely disappearing during the eruption. 
Boxes 3 and 4 are located within a dimmed loop system to the west. While one would expect a similar recovery behavior for these two locations, they show rather different behavior. Box 3 shows a gradual, almost linear recovery which is followed by a steep rise in emission at 21:30 UT due to rapid loop expansion in the core of the active region. 
Box 4 is located further out in the periphery and is not affected by the expansion of these loops showing a rather constant level of intensity.
Box 2 has a drastic decrease in intensity just before the flare peak, followed by a two-step recovery behavior in 24 hours, similar to the one observed for Box 3. 
The movie shows the same loop systems expanding in both regions, which may explain their similarity in the recovery behavior. 
Box 5, located in an elongated dimming region to the north, shows a sharp decrease in intensity until shortly after the flare peak, which is followed first by a slow and very gradual recovery. This changes into a faster increase in intensity after 22:00~UT until the region recovers to the pre-event intensity. 
The associated movie shows enhanced activity in this part of the AR and loop dynamics close to this box, which is probably responsible for this behavior.

\subsection{March 8, 2019}
\label{result_wd}

\begin{figure}[!ht]
\centering
\includegraphics[scale=0.5,trim=0 10 0 50, clip]{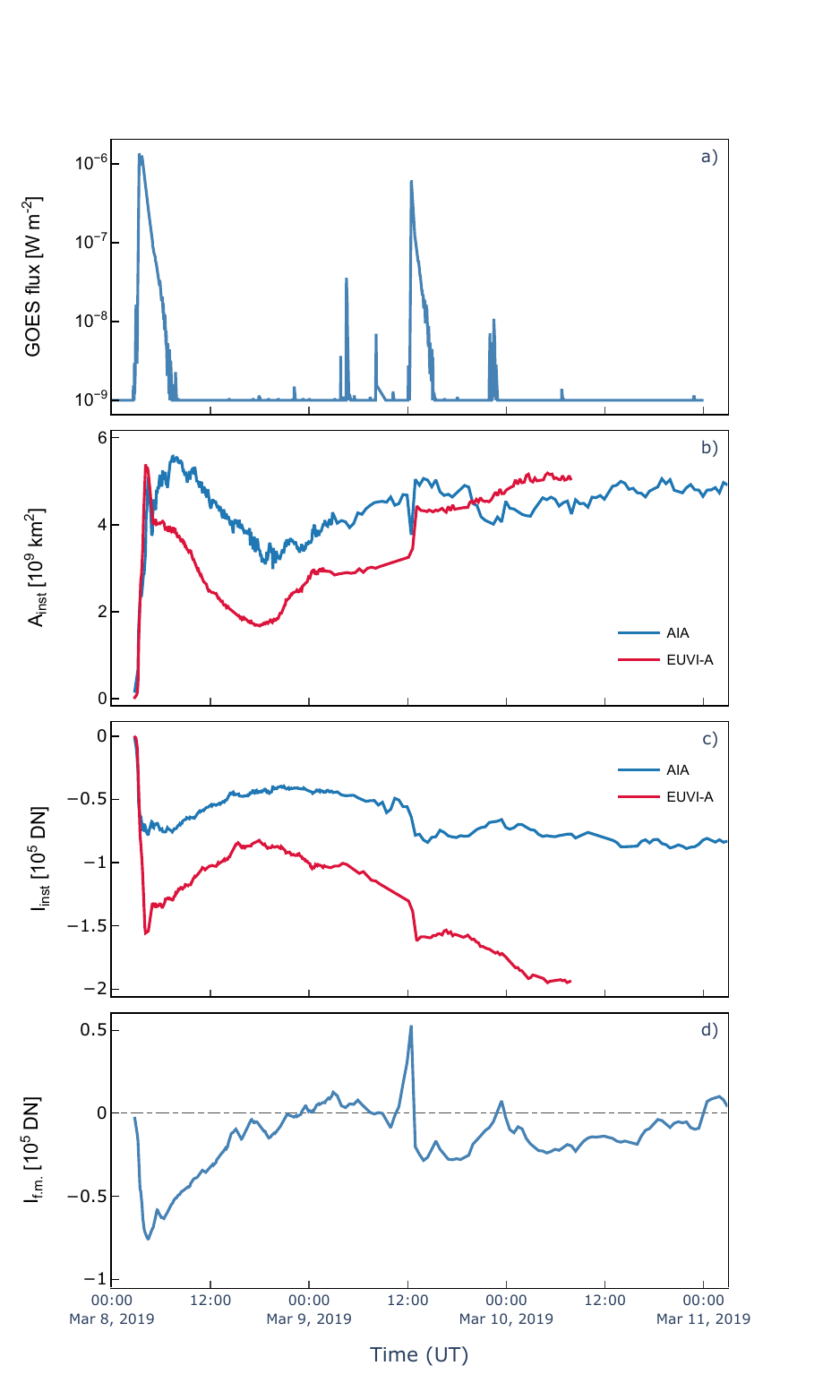}
\caption{Same as in Fig.~\ref{sept-goes}, but for the event on March 8, 2019 (International Women's Day event). For the comparison with STEREO-A data, there is no available information for the third day of the analysis.}
\label{wd-goes}%
\end{figure}

The results for the global dimming recovery for the Women's Day (WD) event on March 8, 2019, are shown in Figure~\ref{wd-goes}. Data from STEREO-A/EUVI were not available for the entire duration of the analysis, but only for approximately 48 hours. For this case, the GOES class is low (C1.0), but the flare is associated with a broad range of phenomena typical for eruptive flares \citep{dumbovic20212019} and the event occurred during solar minimum. 
The instantaneous dimming brightness and area from AIA and EUVI show similar behavior and none of them exhibits a recovery. 
A different situation is shown by the fixed mask brightness curve (Fig.~\ref{wd-goes}d), where the bright emission of the second flare is followed by a decrease in the total brightness, which recovers over a short period (about 17 hours). 
By checking AIA original images in both 211 \AA~and 193 \AA, the expansion of active region loops, similar to the other events, is observed. 

\begin{figure*}[h]
\centering
\resizebox{0.9\hsize}{!}{\includegraphics{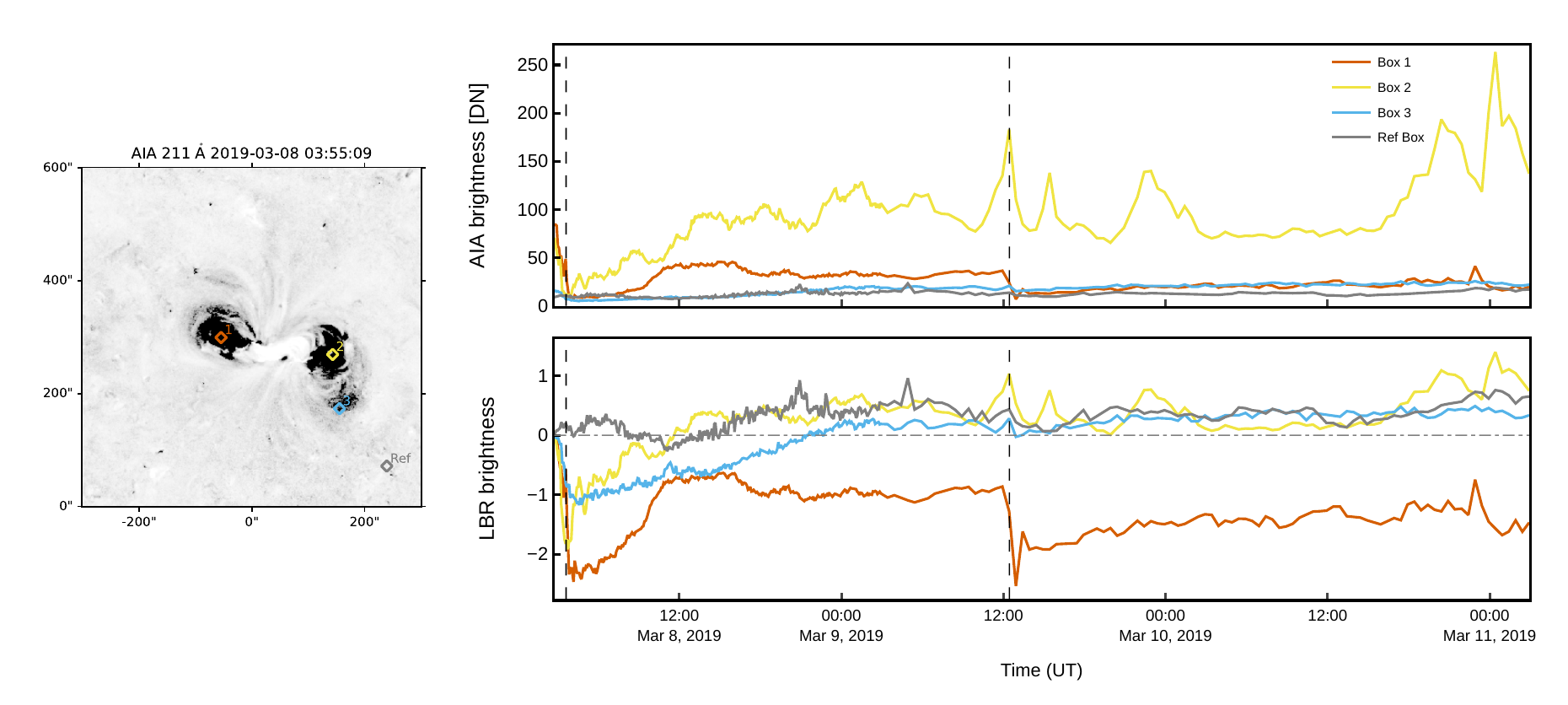}}
\caption{Same as Fig. \ref{im:plot box-sept}, but for the 2019 March 8 event.} 
\label{fig:wd-box-brightness}%
\end{figure*}

Figure~\ref{fig:wd-box-brightness} shows the locations of three pixel boxes on top of a LBR image (left panel) and the time evolution of the average brightness from the original 211 \AA~and LBR data (right panel).
Box 3 shows a linear, gradual recovery behavior and returns to its pre-event intensity within approximately 17 hours. 
On the other hand, box 1 is placed at the center of the eastern region of the dimming and shows a fast recovery (around 8 hours), followed by an almost stationary intensity value until the second flare takes place. 
Box 2 is placed inside the western region of the dimming.  It shows an even steeper intensity decrease as box 1, but a similar recovery rate, which transitions into a constant intensity value after 8 hours. While the intensity of box 1 does not return to the pre-event level, it does for box 2. The similar intensity trends for the recovery of boxes 1 and 2 indicate that they could be core dimmings and therefore part of a the same underlying magnetic structure, while box 3 belongs to the periphery of overlying field and shows a more gradual recovery behavior. We note that the asymmetry in intensity values for boxes 1 and 2  after the dimming recovery could be related to projection effects of loop structures in specific image pixels and solar rotation effects for difference images that become apparent especially for long time series.

\subsection{Comparison of the events}
\label{chap:comparison}

\begin{figure*}[!ht]
\centering
\resizebox{0.8\hsize}{!}
{\includegraphics[width=0.8\textwidth,trim=0 20 10 60, clip]{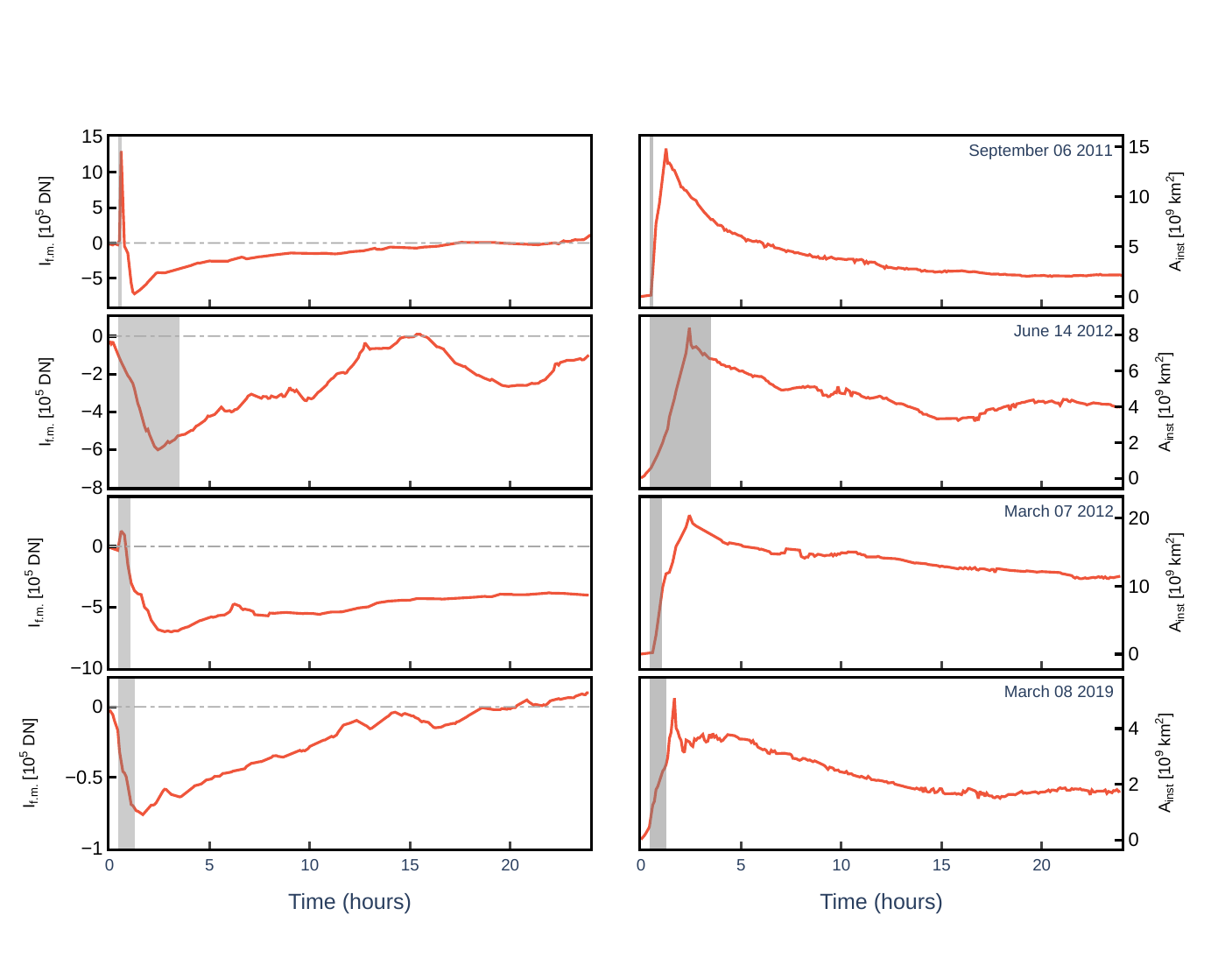}}
\caption{Evolution over 24h of the total brightness of the fixed mask region (left) and of the instantaneous area of the dimming contained within the fixed mask (right) for each of the events. The shaded rectangle marks the GOES flare duration. The dashed horizontal line marks the pre-event brightness level.}
\label{fig: events-comparison}%
\end{figure*}

The four events under study are very different from each other, because of the intensity of the flare, their location on the solar disk, the speed of the associated CME, their duration and their occurrence with respect to the solar cycle. The different sequences of flares over the analyzed time interval also play a part in the diversification of each case. For instance, for September 6, 2011, three flares/CMEs take place within 48 hours, whereas for June 14, 2012, only one long-duration flare, associated with a CME, occurs. 
As a consequence, the intensity and area behavior of the coronal dimming are different for each case under study. 
Figure~\ref{fig: events-comparison} shows for all four events the fixed mask intensity curves in the left column and the instantaneous area evolution of the dimming contained within the fixed mask in the right column. 
The gray rectangles mark the flare time interval and the displayed time is limited to 24 hours to focus on the behavior of the first flare/CME and to better highlight the two-step recovery process in each dataset without the interference of the following eruptions.

In all four cases, the intensity drastically drops when the flare starts. This decrease does not happen in the same way for all the events. The September 6, 2011, and March 7, 2012, events first show an increase in brightness, corresponding to the presence of the bright flare emission inside the fixed mask regions, while June 14, and March 8, 2019, show a small decrease in intensity a bit earlier than the flare start time.
After the minimum intensity value is reached, the recovery starts.
The fixed mask intensity shows a two-step behavior in all four cases, with a steep initial recovery slope followed by a flatter recovery. 
This is particularly evident in the September 6, 2011, case, but this trend is observable in the other three cases, too. 
For the instantaneous dimming area inside the fixed mask, this behavior is less noticeable, apart from the March 8, 2019, case, where a very sharp drop in the area is followed by a milder recovery trend. 
Finally, we also note that the brightness decrease in the Woman's Day event is much smaller than for the other three events, by an order of magnitude.

Out of the four cases under study, three show a recovery to the pre-event level both in brightness and area within 24 hours, while one (March 7, 2012) does not recover to pre-event intensity levels within the analysis time range of 72 hours.

The pixel box approach allows for localized investigation of dimming areas, revealing that observed two-step recovery trend is typically confined to the core dimming regions. The core regions generally show a rapid recovery, plateauing at a constant level, while the gradual complete recovery within 24 hours is attributed to the peripheral dimming regions. 
In some cases, location-specific factors may contribute to variations. 
For instance, in the March 7, 2012, event, a non-recovering region includes peripheral areas, possibly due to significant rotational artifacts. 
In the June 14, 2012, event, the dimming region near a semi-open structure does not recover throughout the observation period. A detailed description of the factors leading to these recovery differences is provided in Section \ref{sec: discussion}.

\section{Summary and discussion}
\label{sec: discussion}

The aim of this work is to elaborate on what has been studied so far about dimming recovery and to implement and test methodologies that could help in better understanding the physical processes behind it.

To this aim, four case studies have been performed. Limitations have been found in terms of observation time for events close to either the western or eastern limbs, because they are subject to distortions and artifacts due to differential rotation. New approaches have been identified to reduce the effect of these challenges, by restricting the analysis to very specific regions, with the application of the maximum dimming area mask to the entire dataset or with the analysis on a smaller scale, with pixel boxes located in regions of interest within the dimming.  The obtained results show similarities with the work of \cite{attrill2008recovery} and \cite{vanninathan2018plasma}, with the identification of core dimmings, showing a slow and sometimes not ending recovery, and secondary dimmings, located further away from the center of the eruption, being the first regions to recover, with an almost linear, gradual trend.

By using the fixed mask approach, the dimming recovery is investigated on a global scale. Figure~\ref{fig: events-comparison} shows that in three events (September 6, 2011; June 14, 2012; and March 8, 2019) the total intensity within the fixed mask region shows a complete recovery within 24 hours, while one event (March 7, 2012) does not recover to pre-event intensity levels within the analysis time range of 72 hours. 
We also observe that the global recovery behavior follows a two-step evolution, with a fast and steeper segment first and a longer and flatter segment later. This trend is observable in all the four cases, but is particularly evident in the September 6, 2011, case. This result is in accordance with \cite{reinard2008coronal}, who
found that statistically about one-fourth of the events showed a two-step evolution in the intensity and/or area recovery curves.
Here we find that the primary mechanism for recovery appears to be the expansion of the coronal loops into the dimming region (see further discussion below). Nevertheless, in the case of the March 7, 2012, event,  brightenings have been observed, which may point at a replenishment of plasma emerging and reconnecting inside the dimming region.  

The pixel box approach investigates dimming recovery locally, and allows us to compare the behavior of different segments inside the dimming region  in terms of their intensity evolution and recovery (Figures~\ref{im:plot box-sept}, \ref{im:plot box-march}, \ref{im:plot box-june}, \ref{fig:wd-box-brightness}).
This approach may be integrated with recent studies and help to identify core dimming regions, for example in \cite{dumbovic20212019} where the authors did not distinguish between core and secondary dimmings, and investigate their recovery time, in accordance with \cite{attrill2008recovery} and \cite{vanninathan2018plasma}, who found that the core dimming regions do not recovery by the end of the studied analysis time interval. 

Below we present a summary of the main observations from this approach  for the four events under study:

\begin{itemize}
    \item The March 8, 2019, event is the only event in our sample that occurred during solar minimum and shows a simpler magnetic configuration of the host active region compared to the other events. We also note that compared to the other three events, it was associated with a small flare (C1), a slow CME (290 km/s) and its dimming region showed a much shallower intensity decrease (by about an order of magnitude, see Fig.~\ref{fig: events-comparison}). It facilitates the observation of two distinct recovery trends within dimming regions. 
    The linear, gradual recovery, spanning almost 17 hours in the periphery, contrasts with the faster two-step recovery process observed in the region closer to the core part of the dimming. In the latter, a rapid recovery occurs within 8 hours, followed by a plateauing of intensity without further increase throughout the 72-hour observation period. This intensity plateau is at a lower level than the pre-event intensity. Despite the AR simplicity, the influence of coronal loops on the intensity recovery of dimmings underscores the complex interplay between different loop structures. These can be considered consistent with the suggestion by \cite{attrill2008recovery}: it was possible to identify a long-lived part within the dimming that did not recover by the end of the analysis interval of 72 hours, meaning that internal recovery is slower than the process occurring at the peripheral regions. 
     
     \item In the case of the June 14, 2012, event, we locate the area boxes in various parts of the complex active region. Notably, within the potential twin core areas of the dimming, we observed a simultaneous occurrence of the local maximum and the local minimum of the LBR brightness. Specifically, the eastern portion exhibited a significant decrease just before the flare, followed by a two-step recovery process spanning 24 hours. Additionally, the dimming part to the southeast from the active region, within the semi-open structure (PFSS extrapolation in Fig.~\ref{fig:pfss}), consistently maintained a minimal intensity level throughout the entire observation period. \cite{2023Ngampoopun} reported that coronal dimmings have the potential to merge with coronal holes, forming a unified structure during the reconnection process. This merging process may lead to the persistence of coronal dimmings for longer than 72 hours, during which the dimming and coronal hole become indistinguishable for detection purposes based on intensity. However, we note that our observations did not reveal any discernible signatures of the reconnection process occurring in this merging region.

     \item The March 7, 2012, event shows similar recovery patterns. The western dimming regions demonstrate complete recoveries within 24 to 36 hours, whereas the eastern ones remain persistently dim throughout the observation period but reach a constant intensity level within 24 hours as well. However, artifacts induced by differential rotation near the solar limb introduce uncertainties. Despite this, gradual recovery in the southwestern periphery region suggests a two-step process, possibly influenced by surrounding brightenings, indicating a reconnection processes (Fig.~\ref{fig: march-bright}).

     \item The September 6, 2011, event presents a very nuanced view of different recovery behaviors and time scales. Rapid (around 12 hrs) as well more gradual, long-term recoveries are observed, all attributed to the expansion of coronal loops into the dimming region or along its line-of-sight but on different time scales. Once again, longer two-step recoveries in certain regions align with prior classifications of core dimming regions, some of which that do not correspond to classical feet of the erupting flux rope but to opened up coronal field through the reconnection process of the initial eruption. Peripherial dimming regions show a linear, gradual and complete recovery.     
\end{itemize}   

In this work, we also present images and videos of the dimming regions evolution in symmetrically scaled colormaps (Figures \ref{shot-sept} - \ref{shot-WD}), to focus not only on the dimmed parts but also on the bright structures. This allows us to resolve spatial fine structures - coronal loops expanding in the active region before and after the CME eruption, which cover dimming regions and result in an increase in their intensity. This observation is in line with the findings of \cite{attrill2008recovery}, which suggested internal brightenings as post-flare loop expansion, although not directly observed. With SDO/AIA we seem to have reached the spatial and temporal resolution to directly observe the full recovery, in terms of its intensity, as coronal loops cover the dimming.

Complementary STEREO/EUVI observations, shown in Figure~\ref{im:snapshots} and the associated movie for, for example, September 6, 2011, depict a system of coronal loops expanding higher than the post-flare loops. These loops persist for a long time, despite being interrupted by consecutive eruptions, and are also observed at later times. \cite{2013Morgan} reported such persisting systems of closed expanding coronal loops, which occur in contrast to twisted flux rope structures and post-flare arcades. Their prolonged existence suggests they are an inherently stable process above regions of flux emergence on the Sun, not merely a build-up to an eruptive event. 
As was reported, after a large CME, the disruption to the expanding loop system is temporary, followed by a renewed period of expansion for around 24 hours.
We observe such high-lying loops for all studied events and time scales of global dimming recovery of the same order for three out of four events. The pixel box analysis supports that the simultaneous recovery trends in specific parts of dimming regions (boxes closely located to the eastern side for the Women's Day and the March 7, 2012, events, northwestern for June 14, 2012) may result from being covered by the same loop systems.

\begin{figure*}[!ht]
\centering
  \resizebox{0.7\hsize}{!}{\includegraphics{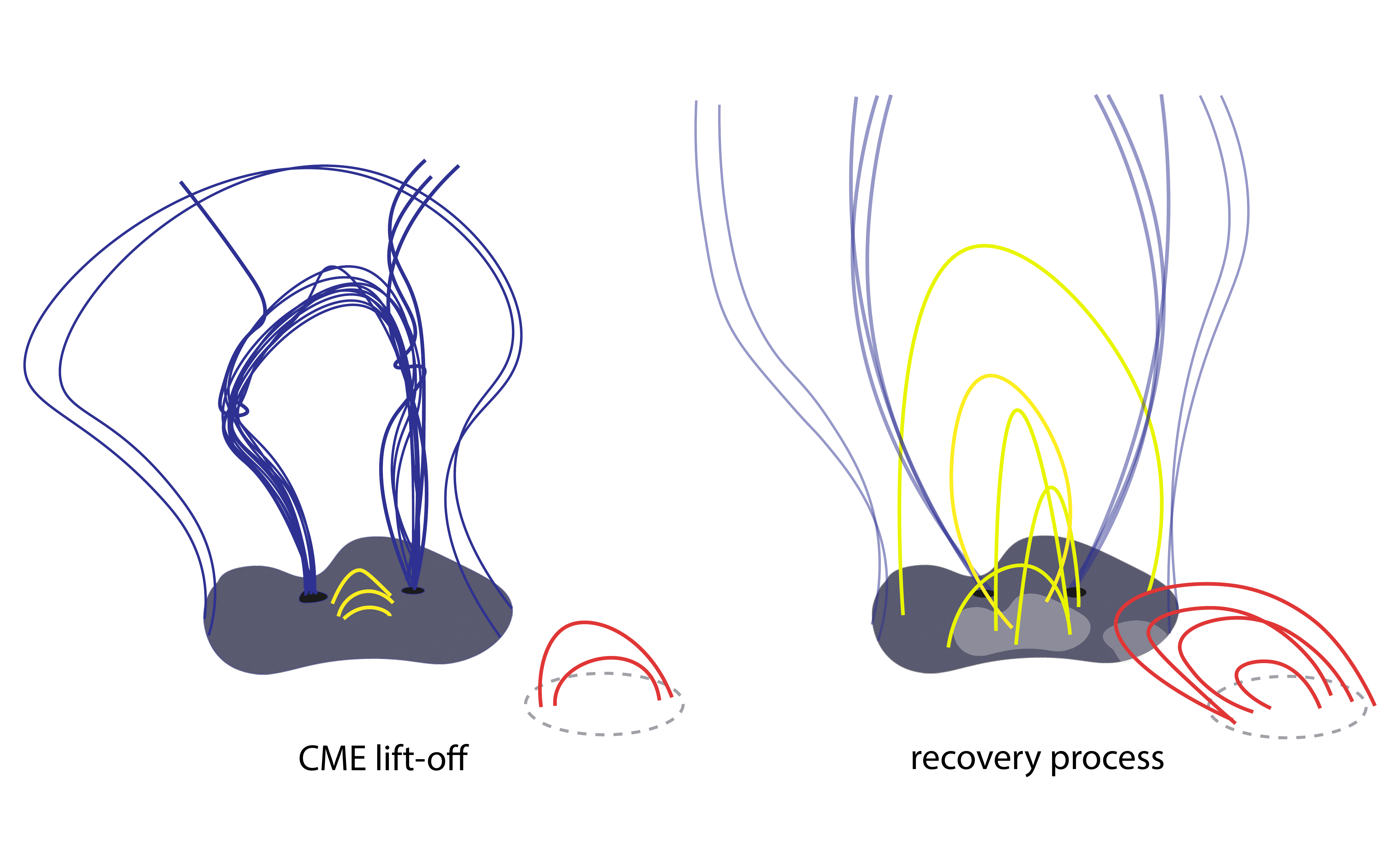}}
  \caption{Cartoon depicting the recovery process due to the expansion of coronal loops (indicated in yellow and red) into the dimming region (gray). The CME structure is shown in blue with dark gray footpoints, referring to core dimming regions. The dashed circle indicates a nearby active region (AR). Recovering areas of the dimming are highlighted in light gray color.
  }
  \label{im:cartoon}
\end{figure*}

A visual representation of the primary dimming recovery mechanism is presented in Figure~\ref{im:cartoon}. The left panel pictures the CME lift-off, in the form of an erupting flux rope and the stretching of an overlying field (in blue), where dimming regions form at its base (gray regions). Loops of the active region that hosts the eruption (yellow) as well as from a nearby active region (red) are also shown. Core dimming regions at the feet of the erupting flux rope are indicated as black patches. The right panel shows the primary recovery process observed: the expansion of the active region, post-flare, and peripheral loops into (yellow) or toward (red) the dimming region that results in an increase in intensity (light gray regions) in these regions.

The observed two-step recovery behavior, with plateaus at lower intensity levels and differing evolution for twin core parts of the dimmings, leads us to question the assumption that the dimming intensity should return to the same pre-event levels. During solar eruptions, the magnetic field undergoes changes, due to magnetic reconnection. Reconnected loops disappear at specific locations and form at new ones, changing the intensity of specific regions over time. Even Sun-as-a-star observations, such as the same March 7, 2012, event reported by \cite{2021Veronig}, show different post-event intensity levels. ARs continuously change their configuration through evolution, making it unlikely to return to its original state.

A systematic observational study combining magnetic field modeling and integrating emission along parts of the loops that are along the line of sight can help to resolve closed and open-field structures and define the true specifics of the dimming recovery.
The comparative analysis of dimming region recovery across multiple events and in different dimming segments underscores the different underlying timescales governing their evolution.
In the future, performing a systematic statistical analysis will be useful to quantify the recovery time of the coronal dimmings by using the fixed mask approach, while the pixel boxes method could be of help to understand the dynamics of different types of dimmings, possibly by integrating the knowledge of already-analyzed events, but focusing more on the recovery phase of the dimming rather than its early evolution.


\begin{acknowledgements}
  G.M.R., G.C. and T.P. acknowledge support by the Russian Science Foundation under the project 23-22-00242, \url{https://rscf.ru/en/project/23-22-00242/}. 
  SDO data are courtesy of the NASA/SDO AIA and HMI science teams.  
  The STEREO/SECCHI data are produced by an international consortium of the Naval Research Laboratory (USA), Lockheed Martin Solar and Astrophysics Lab (USA), NASA Goddard Space Flight Center (USA), Rutherford Appleton Laboratory (UK), University of Birmingham (UK), Max-Planck-Institut für Sonnenforschung (Germany), Centre Spatiale de Liège (Belgium), Institut d’Optique Théorique et Appliquée (France), and Institut d’Astrophysique Spatiale (France). 
  We thank the referee for valuable comments on this study.
\end{acknowledgements}

\clearpage
\bibliographystyle{aa} 
\bibliography{Bibliography} 
\end{document}